\documentclass[12pt]{article}
\pdfoutput=1

\usepackage{epsfig,amsmath,amssymb,verbatim,mathrsfs}

\def\beq{\begin{eqnarray}}
\def\eeq{\end{eqnarray}}
\def\bea{\begin{eqnarray}}
\def\eea{\end{eqnarray}}

\def\gev{\, {\rm GeV}}

\newcommand{\gsim}{\lower.7ex\hbox{$\;\stackrel{\textstyle>}{\sim}\;$}}
\newcommand{\lsim}{\lower.7ex\hbox{$\;\stackrel{\textstyle<}{\sim}\;$}}

\begin{document}

\setlength{\baselineskip}{0.2in}

%#!latex

%\twocolumn[\hsize\textwidth\columnwidth\hsize\csname
%@twocolumnfalse\endcsname
%%
%%
\begin{titlepage}
\noindent
\begin{flushright}
CERN-PH-TH/2007-219 \\
MCTP-07-31  \\
Saclay T07/141
\end{flushright}
\vspace{1cm}

\begin{center}
  \begin{Huge}
    \begin{bf}
    Dynamics of Non-renormalizable \\[.2cm] Electroweak Symmetry Breaking 
     \end{bf}
  \end{Huge}
\end{center}
\vspace{0.2cm}
\begin{center}
\begin{large}
C\'edric Delaunay$^{a,b}$, Christophe Grojean$^{a,b}$, James D. Wells$^{a,c}$ \\
\end{large}
  \vspace{0.3cm}
  \begin{it}
 $^{a}$\, CERN, Theory Division, CH 1211, Geneva 23, Switzerland \\
 $^{b}$\, Service de Physique Th\'eorique, CEA Saclay, F91191, Gif-sur-Yvette, France \\
$^{c}$\, Michigan Center for Theoretical Physics (MCTP) \\
Physics Department, University of Michigan, Ann Arbor, MI 48109

\vspace{0.1cm}
\end{it}

\end{center}

%\center{\today}

\begin{abstract}
\noindent
We compute the complete one-loop finite temperature effective potential for electroweak symmetry breaking in the Standard Model with a Higgs potential supplemented by higher dimensional operators as generated for instance in composite Higgs and Little Higgs models.  We detail the resolution of several issues that arise, such as the cancellation of infrared divergences at higher order and imaginary contributions to the potential.  We follow the dynamics of the phase transition, including the nucleation of bubbles and the effects of supercooling. We characterize the region of parameter space consistent with a strong first-order phase transition which may be relevant to electroweak baryogenesis.  Finally, we investigate the prospects of present and future gravity wave detectors to see the effects of a strong first-order electroweak phase transition.

\end{abstract}

\vspace{1cm}
%\today

\end{titlepage}

\setcounter{page}{2}
%\setcounter{figure}{0}
%\setcounter{table}{0}

%\tableofcontents

\vfill\eject

\section{Introduction}

The baryon asymmetry of the universe remains a mystery. Many ideas have been formulated in the literature, yet much uncertainty remains as to how the
baryon asymmetry could arise.  It is not even clear at what scale the initial asymmetry is produced.
The Sakharov conditions for baryogenesis are baryon number violation, $C$ and $CP$ violation, and
a departure from equilibrium. The Standard Model (SM) does not exhibit these conditions at nearly the strength
required to produce the observed asymmetry given our standard cosmological assumptions, and
thus it is expected that we must go beyond the SM in order to explain the asymmetry.  

The last Sakharov condition,
departure from equilibrium, implies the necessity of a strong first-order phase transition. Since we know that the electroweak symmetry must be broken it is tempting to assume that the corresponding phase transition can satisfy this condition. As noted above, the SM is inadequate, but how far beyond the SM must
one go to find the necessary out of equilibrium dynamics? This question has been addressed by a number of authors (e.g., see~\cite{EW_Phase_Transition_Various_Models} for studies of the dynamics of the electroweak phase transition in various recent models). In Ref.~\cite{Grojean:2004xa} it was shown that
if the  Higgs potential is augmented merely by a $H^6$ operator, it can generate a strong first-order electroweak phase transition.
As one can intuit, the scale suppressing this non-renormalizable operator must be
in the neighborhood of the electroweak scale in order to generate a substantive effect on the
phase transition dynamics. A tree-level analysis 
of this theory was conducted in~\cite{Grojean:2004xa}, with some further refinements 
in~\cite{Bodeker:2004ws}, and it was concluded that a strong first order phase transition is possible even with a Higgs boson as massive as 200~GeV. Of course, for the presence of this $H^6$ operator to be compatible with electroweak (EW) precision data, a higher scale should suppress other dimension six operators, in particular those leading to oblique corrections. The analysis of~\cite{Giudice:2007fh} shows that the low energy effective theories of strongly interacting models, where a light composite Higgs emerges as a pseudo-Goldstone boson,  have precisely
this structure and single out $H^6$ as one of the dominant dimension six operators~\footnote{{In the case of a strongly interacting light Higgs boson, the general effective lagrangian includes four operators that are genuinely sensitive to the strong dynamics~\cite{Giudice:2007fh}, i.e.\ suppressed by the strong decay constant and not the masses of the heavy resonances or the cutoff scale of the strong sector. In this context we can concentrate on the $H^6$ operator since it is the only one that affects the shape of the potential at tree-level, and thus it  has significant effects on the  dynamics of the phase transition as we shall see. }}, being suppressed by the decay constant of the strong sector, parametrically lighter than the cutoff scale of the model. For a fixed value of the strong decay constant, the compatibility with precision EW data is ensured by pushing the masses of vector resonances above 2.5~TeV~\cite{Giudice:2007fh}. In that case a decay constant as low as 300~GeV would be compatible with precision measurements. Our analysis should also apply to study the dynamics of electroweak symmetry breaking in Little Higgs models, and to the more general cases where the $H^6$ operator is generated by integrating out a heavy massive scalar field. However, some extra fine-tunings might be needed in that case to evade EW precision data.

In this publication we extend the results of~\cite{Grojean:2004xa,Bodeker:2004ws} in several ways. First,  we re-analyze the theory using the full finite temperature
effective (nonrenormalizable) Higgs potential at one-loop.   Second, we study the nucleation of broken phase bubbles and consider the effects of supercooling on the electroweak phase transition (EWPT) within this more complete analysis. This is an important dynamical consideration of the phase transition that can in principle have dramatic consequences to when (and if) the phase transition happens. Finally, we investigate whether or not the gravitational waves emitted at the nucleation time can be detected by present and future interferometry experiments, which  would provide another way to study the origin of EW symmetry breaking and another way to test the composite nature of the Higgs.  We consider each of these points in the following three sections,
and then make some concluding remarks.

%%%%%%%%%%%%%%%%%%%%%%
\section{One-Loop Finite Temperature Effective Potential}

Once non-renormalizable interactions are allowed in the theory, as in our case, complete renormalization requires that the infinite set of higher-order operators be considered. However, one is able to truncate the list of needed operators in a perturbative expansion of the inverse cutoff scale.  To study the effect of new physics on the Higgs potential in this effective field theory context, it is sufficient to work at the order $\Lambda^{-2}$ where $\Lambda$ is the cutoff scale suppressing the effective operators. Higher dimensional operators will be sufficiently irrelevant to our problem and can be ignored. 

Our analysis is focussed on operators  that affect the Higgs self-interactions. 
These effective interactions parametrize the new physics responsible for EW symmetry breaking that become fully dynamical at about the scale $\Lambda$. Thus they can be used to generically constrain beyond-the-SM physics affecting the Higgs sector.  Though EW precision measurements put severe constraints on the set of operators affecting the weak bosons' polarization tensors, the effective Higgs self-interactions are almost completely free parameters since the Higgs sector has not yet been probed directly by experiment. Thus the scale suppressing the operator $H^6$ we will focus on can be significantly lower than the cutoff scale of the (strongly coupled) model. This is in particular the case of composite Higgs models when the Higgs emerges from a strongly-interacting sector as a light pseudo-Goldstone boson~\cite{Giudice:2007fh}. The scale suppressing the $H^6$ operator is then $f$, the decay constant of the strong sector, a quantity 4$\pi$ smaller that the cutoff scale.

We start with the following classical effective potential for the SM Higgs \cite{C.Citations}:
\begin{equation}\label{eqn:classpot}
V(H)=m^2|H|^2+\lambda|H|^4+\kappa |H|^6
\end{equation}
where 
$H^T=\left(\chi_1+i\chi_2,\varphi+i\chi_3\right)/\sqrt{2}$ which develops a vacuum expectation value (VEV)  equal to $v_0\simeq 246\gev$.  $\kappa^{-1/2}$ is  identified  at tree level with the decay constant of the strong sector -- the details of this identification at one-loop are described later. We choose a vacuum configuration where only the real part of the neutral component has a constant background value: $\varphi=\phi+h$. The physical Higgs boson is $h$, and we use the traditional background field method~\cite{Jackiw:1974cv} to evaluate the quantum potential for $\phi$ at one-loop. We focus on the main relevant contributions coming from the $SU(2)_L\times U(1)_Y$ gauge bosons, the top quark,  and the Higgs and Goldstone scalars. 

As we briefly review in Appendix~\ref{app:review_technics}, the quantum potential for the background value up to one-loop order at finite temperature in the Landau gauge (where ghosts decouple) is
\begin{equation}\label{eqn:Vtree+1}
V_{eff}(\phi,T)\equiv V_{tree}(\phi)+\Delta V_1(\phi,T)
\end{equation}
with
\begin{eqnarray}
V_{tree}(\phi)&=&\frac{m^2}{2}\phi^2+\frac{\lambda}{4}\phi^4+\frac{\kappa}{8}\phi^6, \\
\Delta V_1(\phi,T)&=&\sum_{i=h,\chi,W,Z,t}\frac{n_iT}{2}\sum_{n=-\infty}^{+\infty}\int \frac{d^3\vec k}{(2\pi)^3}\log\left[\vec{k}^2+\omega_n^2+m^2_i(\phi)\right]
\label{eqn:fullV1eff}
\end{eqnarray}
where $k_E=(\omega_n,\vec{k})$ is the euclidean loop 4-momentum, $\omega_n$ are the Matsubara frequencies in the imaginary time formalism, where $\omega_n=2n\pi T$ for bosons (periodic on the euclidean time circle) and $\omega_n=(2n+1)\pi T$ for fermions (anti-periodic on the euclidean time circle). 
The numbers of degrees of freedom for the relevant fields are $n_{\{h,\chi,W,Z,t\} }=\{1,3,6,3,-12\}$. We include the fermion-loop minus sign in the definition of $n_t$.

Note that in the Landau gauge one must count all three degrees of freedom of each massive vector boson {\it and} the one degree of freedom of each Goldstone scalar. 
This may be qualitatively understood be recalling that the $\chi_i$ Goldstone fields are independent quantum fluctuations away from the zero-temperature minimum. We present a quantitative argument showing this is not double counting in Appendix \ref{app:review_technics}. 
 
We obtain the background-dependent masses appearing in (\ref{eqn:fullV1eff}) by expanding the theory about the background value $\phi$ and reading off the quadratic terms for the various quantum fluctuations. In our dimension-six model the masses are
\begin{eqnarray}
m^2_h(\phi) &=& m^2+3\lambda \phi^2 +\frac{15}{4}\kappa\phi^4, \label{eqn:massh}\\
m^2_\chi(\phi)&= & m^2+\lambda \phi^2 +\frac{3}{4}\kappa\phi^4,\label{eqn:masschi}\\
m^2_W(\phi)&=&\frac{g^2}{4}\phi^2,\ m^2_Z(\phi)\,\,\,=\,\,\,\frac{g^2+g'^2}{4}\phi^2,\ m^2_t(\phi)\,\,\,=\,\,\,\frac{y_t^2}{2}\phi^2,\label{eqn:nonscalarmasses}
\end{eqnarray}
where $g$,$g'$ and $y_t$ are the $SU(2)_L$, $U(1)_Y$ and top Yukawa couplings respectively.
At the zero-temperature minimum one recovers $m^2_h(v_0)=m^2_h$ and $m^2_\chi(v_0)=0$. Note that the expressions for the masses of the weak bosons (from the Higgs kinetic term) and the top quark (from the Yukawa coupling) are unchanged compared to the SM, and (\ref{eqn:nonscalarmasses}) are written to confirm our conventions.

The one-loop correction (\ref{eqn:fullV1eff}) splits into a zero-temperature part and a $T$-dependent part~\cite{Dolan:1973qd,Quiros:1999jp} which vanishes as $T\rightarrow 0$:
\begin{equation}\label{eqn:V1split}
\Delta V_1(\phi,T)\equiv \Delta V_1^0(\phi)+\Delta V_1^T(\phi,T)
\end{equation}
with
\begin{eqnarray}
\Delta V_{1}^0(\phi)&=&\sum_{i=h,\chi,W,Z,t}\frac{n_i}{2}\int \frac{d^4 k_E}{(2\pi)^4}\log\left[k_E^2+m^2_{i}(\phi)\right]\\
\Delta V_1^{T}(\phi,T)&=&\sum_{i=h,\chi,W,Z,t}\frac{n_i T^4}{2\pi^2}\int_0^\infty dk k^2\log\left[1\mp e^{\left(-\sqrt{k^2+m_i^2(\phi)/T^2}\right)}\right]
\end{eqnarray}
$\Delta V_{1}^0(\phi)$ is precisely the ordinary zero temperature effective potential, as it must be to be consistent since $\Delta V_1^T(\phi,T)\to 0$ as $T\to 0$. The $T=0$ part, being UV-divergent, will be considered first in order to properly determine the renormalized parameters of the quantum theory. The finite temperature corrections will be treated afterwards.

%%%%%%%%%%%%%%%%%%%%%%%%%%%%%%%%%%%%%%%%%
\subsection{Zero Temperature Corrections}

At zero temperature the correction (\ref{eqn:fullV1eff}) reduces to the first term of (\ref{eqn:V1split}),
\begin{eqnarray}
\Delta V_1(\phi,T=0)\equiv \Delta V_{1}^0(\phi)&=&
\sum_{i=h,\chi,W,Z,t}n_i\frac{m_{i}^4(\phi)}{64\pi^2}\left[\log\frac{m^2_{i}(\phi)}{\mu^2}-C_i-C_{UV}\right]\label{eqn:dimregpot1}
\end{eqnarray}
which has been regularized in $4-\epsilon$ dimensions, $C_i=5/6$ ($3/2$) for gauge bosons (scalars and fermions) and $C_{UV}\equiv\frac{2}{\epsilon}-\gamma_E+\log 4\pi+{\cal O}(\epsilon)$.

We work in the $\overline{MS}$ scheme to renormalize and evaluate our potential (see the Appendix~\ref{sec:T=0corrections} for an alternative, but ultimately equivalent, on-shell scheme approach).  The full one-loop
effective potential is
\beq
V_{eff}(\phi)=\frac{m^2}{2}\phi^2+\frac{\lambda}{4}\phi^4+\frac{\kappa}{8}\phi^6+\Delta V^0_1(\phi)
\eeq
where the parameters of this potential ($m^2,\lambda,\kappa$) are bare parameters, but an implicit $\delta V_{CT}$  will cancel their infinite pieces, leaving the finite pieces as the renormalized parameters. 

To determine the parameters of the lagrangian in terms of physical quantities, we must impose renormalization conditions at some chosen scale $\mu_*$.  The renormalization conditions are
\bea
V'_{eff}(\phi=v_0,\mu_*) & = & 0 \\
V''_{eff}(\phi=v_0,\mu_*) & = & m_h^2 \\
V'''_{eff}(\phi=v_0,\mu_*) & = & \xi 
\eeq
The left side of each equation is the theory computation, and depends on the parameters of
the theory ($m^2,\lambda,\kappa$). The right side of each equation 
is a measurement ($m_h$ and $\xi$)
or related to a measurement ($V'(v_0)=0$ is a requirement that the potential is at a minimum 
which recovers the correct $Z$ boson mass).  The VEV depends on the choice of scale as well. We define
$v_0$ to be equal to the VEV of the Higgs field in the Landau gauge at $\mu=m_Z$ such that the 
$\overline{MS}$ $Z$ mass is recovered.  Performing our computations with
the latest electroweak precision
measurements~\cite{LEPEWWG}, we find  $v_0=246.8\gev$ to a good approximation for a
Higgs mass in our range of interest ($115\gev<m_h\lsim 300\gev$).  This  Higgs VEV
is close to the $246.2\gev$ value in~\cite{Arason:1991ic}.

We can invert these equations to obtain the theory
parameters as a function of measurements:
\beq
m^2_* & = & m^2(m_h^2,\xi,v_0,\mu_*) \\
\lambda_* & = & \lambda(m_h^2,\xi,v_0,\mu_*) \\
\kappa_* & = & \kappa(m^2_h,\xi,v_0,\mu_*)
\eeq
Note, the parameters have scale dependence, and we have defined $m^2_*\equiv m^2(\mu_*)$, etc.

Up to now we have glossed over some important subtleties.  The physical Higgs mass  must be defined at $p^2=m_h^2$, whereas the one-loop effective potential is constructed for $p=0$. To take account of this, and retain the label $m_h^2$ for the physical Higgs boson mass, we need to rewrite the renormalization condition as
\beq
m^2_h \to m^2_h-\Sigma(m^2_h)+\Sigma(0),
\eeq
where $\Sigma(p^2)$ is the two-point function of the Higgs boson (numerically, we used the LoopTools software~\cite{Hahn:1998yk} to evaluate this two-point function).  This approach has the added benefit that the IR singularity in $V''_{eff}(v_0)$ as the Goldstone mass goes to zero is canceled by the IR singularity in $\Sigma(0)$.  We discuss these IR singularity issues in more detail in the Appendix~\ref{app:onshell}. 

The physical parameter $\xi$ is not a unique choice for how to parametrize the measured tri-Higgs coupling, and we wish to rewrite it in a more convenient manner.  First, like the Higgs mass, the Higgs tri-scalar coupling has IR divergences at $p=0$ when the Goldstone bosons become massless.  These IR divergences are also not dangerous because they are matched by the IR divergences of $V'''_{eff}(v_0)$, and cancel in measured cross-sections.  Thus, it is convenient to separate out this IR divergence when parametrizing the tri-Higgs coupling observable: $\xi\equiv \xi_F+\Gamma_{IR}$,
where $\Gamma_{IR}$ contains IR sensitive Goldstone terms\footnote{Explicitly, $\Gamma_{IR}$ is given by $\Gamma_{IR}=\frac{n_\chi}{32\pi^2}\left[3 m^2_\chi(v_0)'' m^2_\chi(v_0)'\log m^2_\chi(v_0)+\frac{\left[m^2_\chi(v_0)'\right]^3}{m^2_\chi(v_0)}\right]$, where $m^2_\chi(\phi)=m^2_*+\lambda_*\phi^2+\frac{3}{4}\kappa_*\phi^4$. $\Gamma_{IR}^{SM}$ is given by the above expression where the limit $\kappa_*\rightarrow 0$ is taken in $m^2_\chi(\phi)$.}.
Furthermore,  since the tri-Higgs coupling $\xi$ in the SM is fixed once the Higgs mass is known, we would like our convention to reflect this manifestly in the decoupling limit of $\kappa\to 0$,
\beq
\lim_{\kappa\to 0} \xi \to \xi^{SM}\equiv \xi_F^{SM}+\Gamma_{IR}^{SM}
\eeq
For finite values of $\kappa$, the deviations of $\xi_F$ from $\xi^{SM}_F$ can be defined by convention to be 
\beq
\xi_F\equiv \xi^{SM}_F+\frac{6v^3_0}{f^2}
\eeq
This convention (i.e., the factor of 6)
ensures that $\kappa^{-1/2}$ can be identified directly as the decay constant of the strong sector, $f$, at tree level.
Putting these elements together, we can now rewrite the third renormalization condition as
\beq
V'''_{eff}(v_0)=\xi \equiv \xi^{SM}_F+\frac{6v^3_0}{f^2}+\Gamma_{IR}.
\label{eq:xi}
\eeq
We emphasize that eq.~(\ref{eq:xi}) is merely a reparametrization of the tri-Higgs physical observable in terms of the decay constant, $f$, rather than $\xi$ for the benefits described above, and that $\xi^{SM}_F$ is a computable function of $m_h$.  

Following the prescription provided above, all the parameters of our Higgs potential ($m^2,\lambda,\kappa$) can now be written in terms of physical observables ($v_0,m_h,f$).  Thus, we are now able to analyze the potential using physical observables as inputs.

%%%%%%%%%%%%%%%%%%%%%%%%%%%%%%%%%%%%%
\subsection{Finite temperature corrections}

From the splitting~(\ref{eqn:V1split}) of the full one-loop effective potential into a $T=0$ part and a $T\neq 0$ part,  we get that the latter finite temperature component is:
\begin{eqnarray}
\Delta V_1^{T}(\phi,T)&=&\sum_{i=h,\chi,W,Z,t}\frac{n_i T^4}{2\pi^2}\int_0^\infty dk k^2\log\left[1\mp e^{\left(-\sqrt{k^2+m_i^2(\phi)/T^2}\right)}\right]\label{eqn:potTcorrections}\\
&\equiv&\sum_{i={\rm bosons}}\frac{n_iT^4}{2\pi^2} J_b\left(\frac{m^2_i(\phi)}{T^2}\right)+\sum_{i={\rm fermions}}\frac{n_iT^4}{2\pi^2} J_f\left(\frac{m^2_i(\phi)}{T^2}\right)\nonumber
\end{eqnarray}
where the upper (lower) sign stands for bosons (fermions). In the high-temperature regime ($T \gg m_i(\phi)$), the $J_i$ function expansions are
\begin{eqnarray}
J_b\left(x\right)&\underset{x\rightarrow 0}{=}&\frac{\pi^2}{12}x-\frac{\pi}{6}x^{3/2}-\frac{x^2}{32}\log\frac{x}{a_b}+{\cal O}\left(x^3\log\frac{x^{3/2}}{\mbox{cst.}}\right)\label{eqn:highTb}\\
J_f\left(x\right)&\underset{x\rightarrow 0}{=}& -\frac{\pi^2}{24}x-\frac{x^2}{32}\log\frac{x}{a_f}+
{\cal O}\left(x^3\log\frac{x^{3/2}}{\mbox{cst.}}\right)\label{eqn:highTf}
\end{eqnarray}
with $\log a_b\simeq 5.4076$ and $\log a_f\simeq 2.6350$. Note that in \cite{Grojean:2004xa} only the first terms in (\ref{eqn:highTb}) and (\ref{eqn:highTf}) were retained, which leads to the following approximate thermal one-loop correction:
\begin{equation}
\Delta V_{1,GSW}^T(\phi,T)\equiv\sum_{i={\rm bosons}}\frac{n_iT^2 m^2_i(\phi)}{24}+\sum_{i={\rm fermions}}\frac{n_f T^2 m^2_i(\phi)}{48}\simeq \frac{1}{2}cT^2\phi^2 + \cdots \label{eqn:V1TGSW},
\end{equation}
with $c=(4 m_h^2/v_0^2+3g^2+g'^2+4y_t^2-12 v_0^2/f^2)/16$.

\begin{figure}[t]
	\centering
	\includegraphics[width=0.60\hsize]{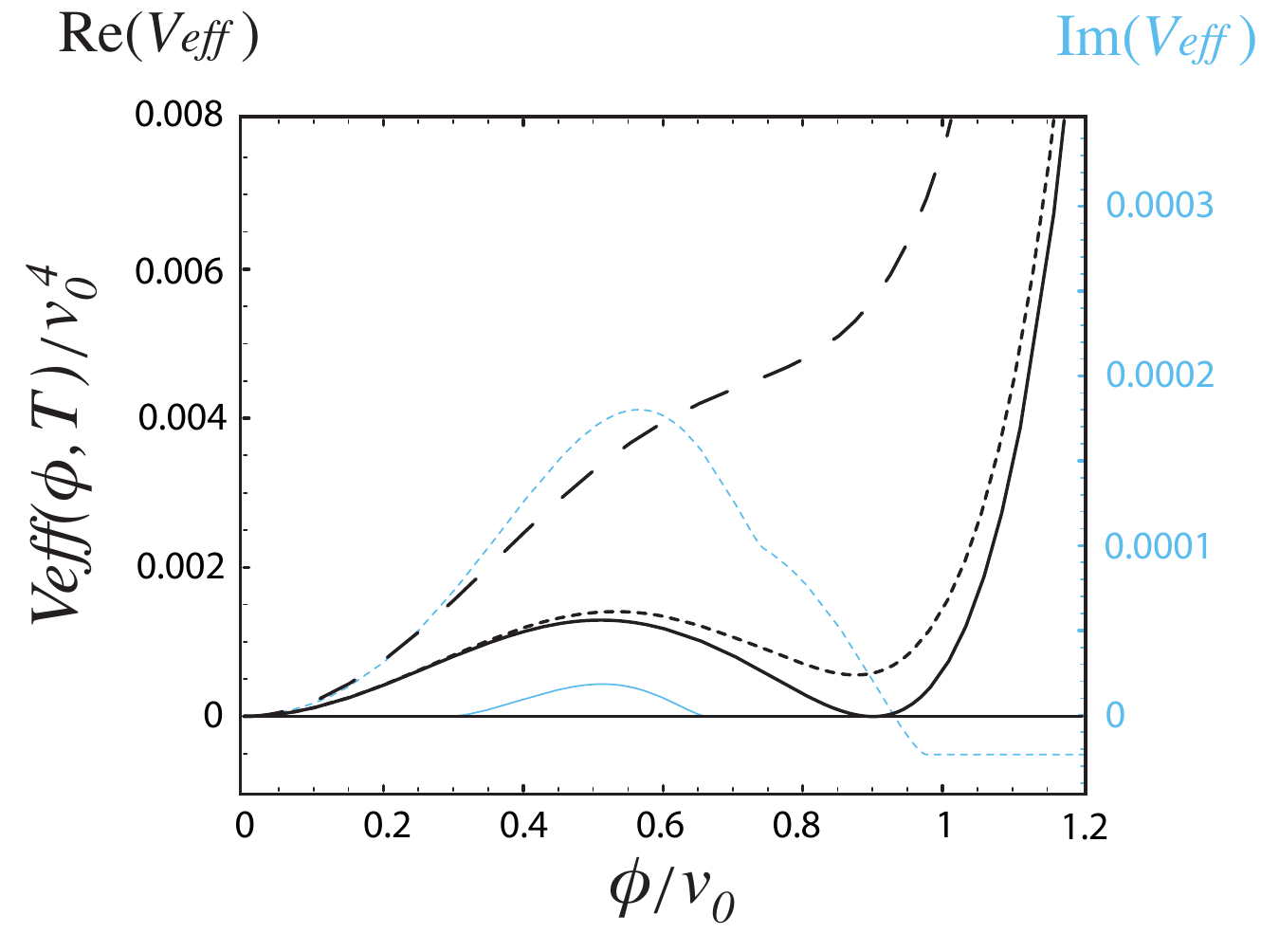}
	\caption{Different potentials close to the critical temperature for $m_h=115$~GeV and $f=620$~GeV ($f$ is the decay constant of the strong sector the Higgs emerges from). The dashed curve is the potential of \cite{Grojean:2004xa} which includes only the  thermal mass term of the Higgs, while the solid and dotted ones represent the full one-loop potential with (solid) and without (dotted) the ring diagram contributions. In blue, we have also plotted the imaginary part of the full one-loop potential with the ring contributions (solid blue) as well as the imaginary part of the ring contributions alone (dashed blue). This illustrates the cancelation of the large imaginary parts between the ring and the one-loop contributions, while there still exists an additional and  smaller imaginary part for some values of $\phi$ due to a negative quartic coupling (see the discussion in Section~\ref{sec:ImCancelation} for details). An imaginary part of the potential can be interpreted as a decay rate of some quantum states of the scalar fields to some others but the imaginary part of the full potential is always tiny compared to the real part around the transition temperature and the system is stable enough throughout the entire time of the transition.}
	\label{fig:Potentials}
\end{figure}

The dominant contributions gathered in (\ref{eqn:V1TGSW}) are simply a (positive) thermal mass which (meta)-stabilizes the origin of the potential at high temperature. This approximation was sufficient in \cite{Grojean:2004xa}, and further refined in~\cite{Bodeker:2004ws}, to demonstrate the possibility of a strong first order PT within an effective extension of the SM.
Fig. (\ref{fig:Potentials}) shows the discrepancy between the complete thermal correction and the high-temperature expansion around the critical temperature, illustrating the worthwhileness of using the integrals of (\ref{eqn:potTcorrections}) for the more detailed analysis. 

%%%%%%%%%%%%%%%%%%%%%%%%%%%%%%%%%%%%%%%%%
\subsubsection{Breakdown of perturbation theory and ring diagrams}

\begin{figure}[t]
	\centering
	\includegraphics[scale=0.5]{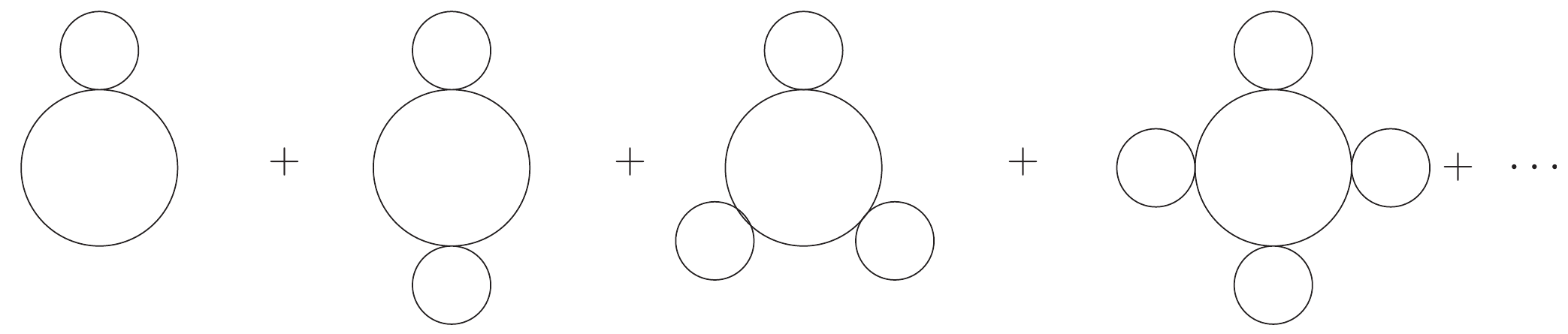}
	\caption{Some generic examples of ring diagrams where each solid line may represent either a scalar, a fermion or a gauge field. The small loops correspond to thermal loops in the IR limit. They are all separately IR divergent, but their sum is IR finite. }
	\label{fig:ring}
\end{figure}

In thermal quantum field theory, the traditional perturbative expansion in terms of small coupling constants 
breaks down due to IR-divergences (inherent in massless models) generated by long-range fluctuations appearing as soon as one moves to finite temperature \cite{Takahashi:1985vx}. For instance, taking  massless $\lambda\phi^4$ theory at finite temperature, one can show that the self-energy, which goes like $\lambda$ at first order, receives a subleading  $\lambda^{3/2}$ correction and not $\lambda^2$ as one would expect~\cite{LeBellac}. For our case, in the high-temperature expansion, or equivalently small mass expansion, of the  thermal bosonic corrections (\ref{eqn:highTb}), we also see a sign of this perturbation theory breakdown through the emergence of a monomial term of order $3/2$. The main consequence is that, as it stands, we cannot trust the completeness of the one-loop result (\ref{eqn:potTcorrections}) because there are some higher-loop corrections of the same order \cite{Dolan:1973qd}, as if the effect of temperature is to ``dilute" the one-loop correction to some multi-loop orders in the IR. Furthermore the leading part of these multi-loop corrections is all contained in the so-called ring (or daisy) diagrams shown in Fig.~(\ref{fig:ring}). They are $N$-loop diagrams where $N-1$ of them are ``ring attached" to a main one. Since this ``loop-dilution" is a finite temperature effect, the ring diagrams only need to be resummed in the IR-limit of vanishing momenta running in their petals \cite{Dolan:1973qd}. It is also well-known that they can be taken into account by using propagators resummed in the IR~\cite{Carrington:1991hz}. By solving a Dyson-like equation, this turns out to simply shift the bosonic masses by a $T$-dependent constant as $m_b^2(\phi)\rightarrow m^2_b(\phi)+\Pi_b(T)$, where $\Pi_b(T)$ is the self-energy of the (bosonic) field $b$ in the IR limit, $\omega=\vec{p}=0$, known as a Debye mass ($\Pi_b(T)$ is labeled as $\Pi_b(0)$ in~\cite{Carrington:1991hz}).

The higher-loop ring diagrams are needed due to IR divergences (i.e., $m\lsim T$). On the other hand,
the one-loop result is trustworthy for massive (i.e., $m\gtrsim T$) particles, because the long-range fluctuations arising at finite temperature will never hit an IR mass-pole in such cases. Hence the ring diagrams will only contribute significantly at high-temperature ($T/m\rightarrow\infty$) where the particles can be approximated as nearly massless. 
Also, this allows us to understand why only the bosonic degrees of freedom feel the breakdown of the perturbative expansion\footnote{In the gauge sector, only the longitudinal polarizations demonstrate this same breakdown of perturbation theory~\cite{Carrington:1991hz}.}. The reason is that only bosonic fields have a vanishing Matsubara frequency, recalling that $\omega_n$ equals $2\pi nT$ for bosons and $(2n+1)\pi T$ for fermions. Only this particular (zero-)mode will behave as a massless degree of freedom and generate IR-divergences at high-temperature, while the other (non zero-)modes $\omega_n$ act as a mass of order $T$ and thus lead to negligible contributions. Therefore the fermionic propagators need not be resummed, because fermions do not have pole-mass in the IR. 

Applying the techniques of~\cite{Carrington:1991hz} to our theory, we compute the finite temperature mass shifts (Debye masses) that are needed in the ring diagram resummation:
\begin{eqnarray}
\Pi_{h,\chi}(T)&=&\frac{T^2}{4v_0^2}\left(m_h^2+2m_W^2+m_Z^2+2m_t^2\right)-\frac{3T^2}{4}\frac{v_0^2}{f^2}\label{eqn:Pih}\\
\Pi_W(T)&=&\frac{22}{3}\frac{m^2_W}{v_0^2}T^2\\ 
\Pi_Z(T)&=&\frac{22}{3}\frac{(m^2_Z-m^2_W)}{v_0^2}T^2-m^2_W(\phi)\\
\Pi_\gamma(T)&=&m^2_W(\phi)+\frac{22}{3}\frac{m^2_W}{v_0^2}T^2.
\end{eqnarray}
Note that these $\Pi(T)$'s are computed in the high-temperature limit of the unbroken phase which is justified by the ring diagrams being irrelevant for $T\lesssim m_i(\phi)$ as we have discussed. At high temperature the photon and $Z$ are not mass eigenstates, but  one can treat them as mass eigenstates in this computation with the above-given Debye masses and obtain the correct resummed potential.

%%%%%%%%%%%%%%%%%%%%%%%%%%%%%%%%%
\subsubsection{Incorporating the ring corrections}
The traditional way the ring diagrams are implemented in the literature consists in shifting all the Matsubara modes for the bosonic fields. This is the so-called self-consistent method \cite{Arnold:1992rz} where the potential (\ref{eqn:fullV1eff}) is replaced by
\begin{equation}\label{eqn:V1Tselfconsistent}
\Delta V_{1+ring}^{self-con.}(\phi,T)=\sum_{i=h,\chi,W,Z,\gamma,t}\frac{n_iT}{2}\sum_{n=-\infty}^{+\infty}\int \frac{d^3k}{(2\pi)^3}\log\left[\vec{k}^2+\omega_n^2+m^2_i(\phi)+\Pi_i(T)\right].
\end{equation}
The thermal shift of the gauge masses only for the longitudinal polarizations is understood, and $\Pi_t(T)$ is simply zero. However, when applying this approach the UV divergent part becomes $T$-dependent through the $\Pi(T)$ and requires $T$-dependent counter-terms to be made finite. Indeed after doing to (\ref{eqn:V1Tselfconsistent}) the same splitting procedure we did to get (\ref{eqn:V1split}), and after dimensionally regularizing the UV-divergent part, we get the following result:
\begin{equation}
\Delta V_{1+ring}^{0,self-con.}=\sum_{i=h,\chi,W,Z,t}n_i\frac{\left(m_{i}^2(\phi)+\Pi_i(T)\right)^2}{64\pi^2}\left[\log\frac{m^2_{i}(\phi)+\Pi_i(T)}{\mu^2}-C_i-C_{UV}\right]
\end{equation}
where the $C_{UV}$ factor depends on $T$. This standard technique clashes with physical intuition since it would mean that the UV behavior of the theory depends on the IR dynamics. Although this mixing is not introducing any calculational errors to our working approximation, one can avoid it by simply shifting only the $\omega_n=0$ Matsubara modes which carry the leading contribution from the ring diagrams relevant at one-loop order. 

As argued above, the dilution of the one-loop correction happens only for massless modes. Hence all the corrections we seek within the ring diagrams are gathered when resumming only the zero-mode of the propagator in the IR. Doing so, (\ref{eqn:fullV1eff}) is to be replaced by
\begin{eqnarray}
\Delta V_{1+ring}(\phi,T)&=&\sum_{i=h,\chi,W,Z,\gamma,t}\frac{n_iT}{2}\Big\{\mathop{\sum\mathstrut'}_{n=-\infty}^{+\infty}\int \frac{d^3k}{(2\pi)^3}\log\left[\vec{k}^2+\omega_n^2+m^2_i(\phi)\right]\nonumber\\
&&+ \int \frac{d^3k}{(2\pi)^3}\log\left[\vec{k}^2+m^2_i(\phi)+\Pi_i(T)\right]\Big\}\\
&\equiv&\Delta V_1(\phi,T)+\Delta V_{ring}(\phi,T)
\end{eqnarray}
where the prime means that the zero modes are excluded from the sum. We can easily extract the ring part from the last expression and we find
\begin{eqnarray}
\Delta V_{ring}(\phi,T)&=&\sum_{i=h,\chi,W,Z,\gamma}\frac{\bar{n}_iT}{4\pi^2}\int_0^\infty dk k^2\log\left[1+\frac{\Pi_i(T)}{k^2+m^2_i(\phi)}\right]\nonumber\\
&=&\sum_{i=h,\chi,W,Z,\gamma}\frac{\bar{n}_iT}{12\pi}\left[m^3_i(\phi)-\left(m^2_i(\phi)+\Pi_i(T)\right)^{3/2}\right]\label{eqn:Vring},
\end{eqnarray}
where an irrelevant (infinite) constant has been ignored in the second line, and $\bar{n}_{\{h,\chi,W,Z,\gamma\}}={\{1,3,2,1,1\}}$. Notice that $\Delta V_{ring}$ includes a monomial of order $3/2$ which proves a posteriori the existence of a perturbation theory breakdown in evaluating the Higgs potential. Furthermore, these extra corrections modify the cubic term in $m_i(\phi)$, which partly controls\footnote{The negative quartic coupling, of course, is another source of a potential barrier for the first-order phase transition.} the strength of the first order phase transition. Thus, the addition of these terms is critical for our analysis of the electroweak phase transition. 

In summary, the full $T$-dependent renormalized effective potential at one-loop is
\bea
V_{eff}(\phi)&=&\frac{m_*^2}{2}\phi^2+\frac{\lambda_*}{4}\phi^4
           +\frac{\kappa_*}{8}\phi^6 + \sum_{i=h,\chi,W,Z,t}n_i\frac{m_{i}^4(\phi)}{64\pi^2}\left[\log\frac{m^2_{i}(\phi)}{\mu_*^2}-C_i\right]\label{eqn:potMSbar} \nonumber \\
 & & +\sum_{i={\rm bosons}}\frac{n_iT^4}{2\pi^2} J_b\left(\frac{m^2_i(\phi)}{T^2}\right)+\sum_{i={\rm fermions}}\frac{n_iT^4}{2\pi^2} J_f\left(\frac{m^2_i(\phi)}{T^2}\right) \nonumber \\
 & & +\sum_{i=h,\chi,W,Z,\gamma}\frac{\bar{n}_iT}{12\pi}\left[m^3_i(\phi)-\left(m^2_i(\phi)+\Pi_i(T)\right)^{3/2}\right]
\eea
where definitions of all terms are given above.
This is the potential we analyze for the remainder  of the article.

%%%%%%%%%%%%%%%%%%%%%%%%%%%%%%%%%
\subsection{Reality of the quantum potential}

As the scalar masses become negative, the various contributions we obtained for the quantum potential develop some imaginary parts 
which we discuss below for both the $T=0$ and $T\neq 0$ cases.

\subsubsection{Imaginary part at $T=0$}

In the zero-temperature limit, the logarithm of (\ref{eqn:potMSbar}) leads to the following scheme-independent imaginary part\footnote{On the principal sheet, the imaginary part of the logarithm is taken to satisfy $ -\pi < \Im m \log \leq \pi$.}
\begin{equation}\label{eqn:IMpart0}
\Im m\left[\Delta V_{1}^{0}(\phi)\right]=\sum_{i=h,\chi}\Theta(-m^2_i(\phi))\frac{n_i |m_i(\phi)|^4}{64\pi}
\end{equation}
where $\Theta(-m^2_i(\phi))$ is the Heaviside function which equals 1 when the field $i$ is tachyonic, and zero otherwise. The Higgs boson
can obtain a negative mass squared for some values of its VEV, originating
from the fact that the classical potential is not convex everywhere. Indeed, depending on the cutoff value, either the origin is unstable  ($f^2>3v_0^4/2m_h^2$) or a potential barrier separates two local  minima ($f^2<3v^4_0/2m_h^2$), both of which lead to concave regions of the effective potential as a function of the VEV. A similar analysis shows that the Goldstone boson can become tachyonic for some values of the VEV as well, leading to another contribution to the imaginary part of the effective potential.  
However, we shall see shortly that the imaginary part (\ref{eqn:IMpart0}) exactly cancels out with another contribution coming from the finite temperature corrections for the temperature range we are interested in for the phase transition.

%%%%%%%%%%%%%%%%%%%%%%%%%%%%%%%%%%%%%%%%%%
\subsubsection{Imaginary part at $T\neq 0$}
\label{sec:ImCancelation}

At finite temperature both the integrals of (\ref{eqn:potTcorrections}) and the ring contributions (\ref{eqn:Vring}) are spoiled by imaginary parts when scalar fields are tachyons. In the high-temperature limit, the imaginary part of (\ref{eqn:potTcorrections}) is (see (\ref{eqn:highTb})):
\begin{equation}
\Im m\left[\Delta V_1^{T}(\phi,T)\right]\underset{\frac{|m_i(\phi)|}{T}\to 0}{\longrightarrow}\  \sum_{i=h,\chi}\Theta(-m^2_i(\phi))n_i\left[-\frac{|m_i(\phi)|^4}{64\pi}+\frac{|m_i(\phi)|^3 T}{12\pi}\right].\label{eqn:IMpartHighT}
\end{equation}
The first term cancels the imaginary part from the logarithm of the $T=0$ potential correction (\ref{eqn:IMpart0}), while the second is only compensated when the ring diagrams are added, since their imaginary part is given by
\begin{equation}\label{eqn:IMpartRING}
\Im m\left[\Delta V_{ring}(\phi,T)\right]=-\sum_{i=h,\chi}\Theta(-m^2_i(\phi))\frac{n_iT}{12\pi}|m_i(\phi)|^3
\end{equation}
as long as the temperature satisfies $m^2_{i}(\phi)+\Pi_{i}(T)>0$ for all $\phi$. 
Although somewhat more complicated algebraically to show (see Appendix~\ref{app:ImCancelation} for details), this cancellation occurs also for smaller temperatures of order $T\sim |m_i(\phi)|$.  

Nevertheless and despite this cancelation, the potential is not everywhere real because for some values of $T$ and $\phi^2$, $m^2_{i}(\phi)+\Pi_{i}(T)<0$ and the second term of the ring correction~(\ref{eqn:Vring}) becomes imaginary.  In the SM this term does not lead to an  imaginary part once the temperature (meta)stabilizes the origin since  the SM scalars could only become tachyonic for a negative quadratic coupling in the Higgs potential. Thus, the SM potential is real as long as the origin is (meta)stable. On the other hand, with the additional $H^6$ piece in the potential, the scalar masses can be negative also through a negative quartic coupling, allowing this additional imaginary part to the potential at temperature around the critical temperature.

An imaginary part of the potential can be interpreted as a decay rate of some quantum states of the scalar fields to some others~\cite{Weinberg:1987vp}. Thus, one can rely on the real part of the potential as long as its imaginary part remains small enough to consider the field stable during the phase transition, in which case it can be discarded.
We checked that the imaginary part of the one-loop potential is always tiny compared to the real part around the transition temperature, thanks to the previously demonstrated cancelations of large imaginary pieces. Thus, we conclude that the system is stable enough throughout the entire time of the transition, and that its dynamics is driven by the real part of the one-loop potential we computed.

%%%%%%%%%%%%%%%%%%%%%%%%%%%%%%%%%%%%%%%%%%%
\section{Dynamics of the Electroweak Phase Transition}

Now that we have the formalism developed for our analysis of the finite temperature Higgs 
potential at one loop, we are in the position to study the dynamics of the phase transition.
One of our first considerations must be the analysis of when (and if) the phase transition
actually occurs. This is not simply a matter of determining the temperature at which the
symmetry breaking minimum becomes the global minimum. An analysis of the energetics
of bubble formation must be undertaken for a more complete picture. The nucleated bubbles can then undergo collisions and the surrounding plasma experience turbulence, which generate gravity waves that could possibly
be detected in experiments. We discuss these issues in this section.

Throughout this section, we report our numerical results of various relevant quantities as contour plots that scan the allowed region of the parameter space $(m_h,f)$. We recall that $m_h$ is the physical Higgs mass while $f$ is the decay constant of the strong sector (or more generally the energy scale suppressing the $H^6$ operator) physically defined through the triple Higgs self-interaction  as defined in the previous sections, and we work in the $\overline{MS}$ scheme for $\mu=m_Z$. The bounds delinating the region of first-order phase transition are both numerically computed using the complete one-loop potential at finite temperature. The lower one is set by requiring that EW symmetry is broken at $T=0$ and restores at high temperature, while above the upper bound the Higgs vacuum is likely to undergo a second-order phase transition or a smooth crossover. In general, determining the latter is not an easy task as it requires a non-perturbative analysis of the effective potential when the transition is not strongly first-order \cite{Non-perturbative_PT}. Indeed, the phase transition always appears first-order at the perturbative level, even though very weakly. Moreover, as $f$ increases one tends to recover the SM potential, which leads non-perturbatively to a continuous crossover, instead of a weak first-order transition at one-loop, for $m_h\gtrsim 80$ GeV \cite{Kajantie:1996mn}. We estimated the upper bound by considering that as soon as the phase transition is as weak as in the SM for $m_h=80$ GeV, it is likely to be a crossover.

%%%%%%%%%%%%%%%%%%%%%%%%%%%%%%%%%%%
\subsection{The onset of nucleation and EW baryogenesis}

The effective potential ensures the presence of a potential barrier at finite temperature which is a necessary ingredient to have a first-order phase transition. It proceeds by spontaneous nucleation of non-vanishing VEV bubbles into a surrounding symmetric metastable vacuum. As soon as the universe cools down to a critical temperature $T_c$ the symmetry-breaking vacuum becomes energetically favorable and then thermal fluctuations allow the bubbles to form. However, the temperature of the  transition is not necessarily close to $T_c$. Once created, a bubble needs to consume a part of the latent heat liberated in order to maintain its interface with the symmetric phase surrounding it. It turns out that for $T$ just below $T_c$ it is often the case that the bubbles are too small and surface tension makes them collapse and disappear. Hence the phase transition effectively starts at a smaller temperature when enough free energy is available to permit the nucleation of sufficiently large bubbles that can grow and convert the entire universe into the broken phase. This supercooling phenomenon can substantially delay the phase transition and thus modify the spectrum of gravity waves significantly, as we shall discuss shortly (important supercooling effects were also observed in some of the analyses of Ref.~\cite{EW_Phase_Transition_Various_Models})

%%%%%%%%%%%%%%%%%%%%%%%%%%%%%%%%%%%%%%
\subsubsection{When does the nucleation start?}

Although the probability to tunnel via the excitation of $SU(2)$ instantons is very tiny, about 
$\exp(-{\cal O}(100))$, the decay of the false vacuum can nonetheless proceed through thermal fluctuations which help to overcome the potential barrier. The rate per unit of space-time for this process is given in the semi-classical WKB approximation by $\Gamma\sim e^{-S_E}$ where $S_E$ is the euclidean action for the Higgs VEV evaluated on the so-called bounce solution of the euclidean equation of motion \cite{Coleman:1977py}. For temperatures much higher than their inverse radius, the bubbles overlap in euclidean time and feel the IR breaking of Lorentz symmetry \cite{Linde:1981zj, Linde:1980tt}, in which case the bounce solution is $O(3)$-symmetric and is the solution of
\begin{equation}
\frac{d^2 \phi_b}{dr^2}+\frac{2}{r}\frac{d \phi_b}{dr}+\frac{\partial V(\phi_b,T)}{\partial\phi_b}=0,
\end{equation}
subject to the boundary conditions
\begin{equation}
\phi_b (r\rightarrow\infty)=0 \mbox{\ \ and\ \ } \frac{d\phi_b(r=0)}{dr}=0. 
\end{equation}
The bounce solution physically represents the Higgs VEV profile of a static unstable (either expanding or shrinking) bubble, and $r$ measures the distance from the bubble center. For such a static solution of the equation of motion, the action factorizes as $S_E=S_3/T$, with
\begin{equation}
S_3=\int dr 4\pi r^2\left[\frac{1}{2}\left(\frac{d\phi_b}{dr}\right)^2+V(\phi_b,T)\right].
\end{equation}
Moreover for small temperatures of the order of the bubble size, we replace the $O(3)$ bounce for the $O(4)$-symmetric solution which minimizes the action when the breaking of Lorentz symmetry is not significant. Finally we use the traditional overshooting/undershooting method to numerically solve the equation of motion.

There is a supercooling effect that can delay the onset of the first order phase transition to temperatures much smaller than $100\gev$. A first order phase transition can only proceed in the presence of a potential barrier separating the two vacua and the nucleation could potentially start at a temperature $T_n$ far below that of $T_c$. This is especially likely in the case where the barrier persists down to $T=0$.
Since the amount of supercooling is controlled by the size of the nucleated bubble, one needs to take into account that the phase transition proceeds in an expanding universe. One can thus consider that the nucleation starts at the time when the probability of creating at least one bubble per horizon volume is of order one. This condition guarantees the percolation of bubbles in the early universe and translates into the following criterion for determining the nucleation temperature:
\begin{equation}\label{eqn:Nucleation}
\frac{S_3(T_n)}{T_n}\sim-4\log\left(\frac{T_n}{m_{Pl}}\right)~ \Longrightarrow~ \frac{S_3(T_n)}{T_n}\sim 
{\cal O}(130-140)\mbox{\ for $T_n\sim 100\gev$}.
\end{equation}
where $m_{Pl}\equiv M_{Pl}/\sqrt{8\pi}$ is the reduced Planck mass. 

\begin{figure}[t]
	\begin{center}
		\begin{tabular}{cc}
	\includegraphics[width=0.48\hsize]{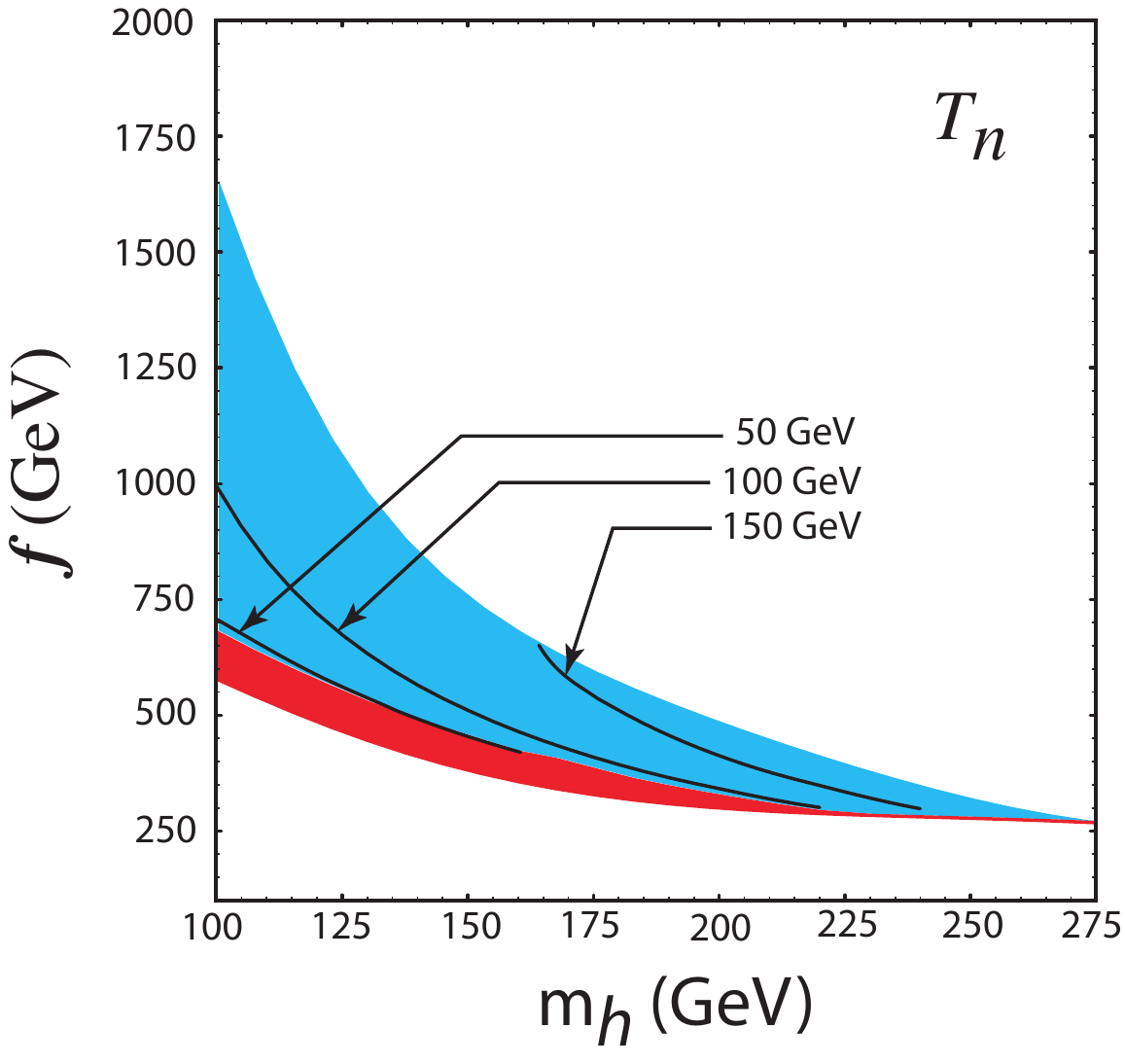}&\includegraphics[width=0.48\hsize]{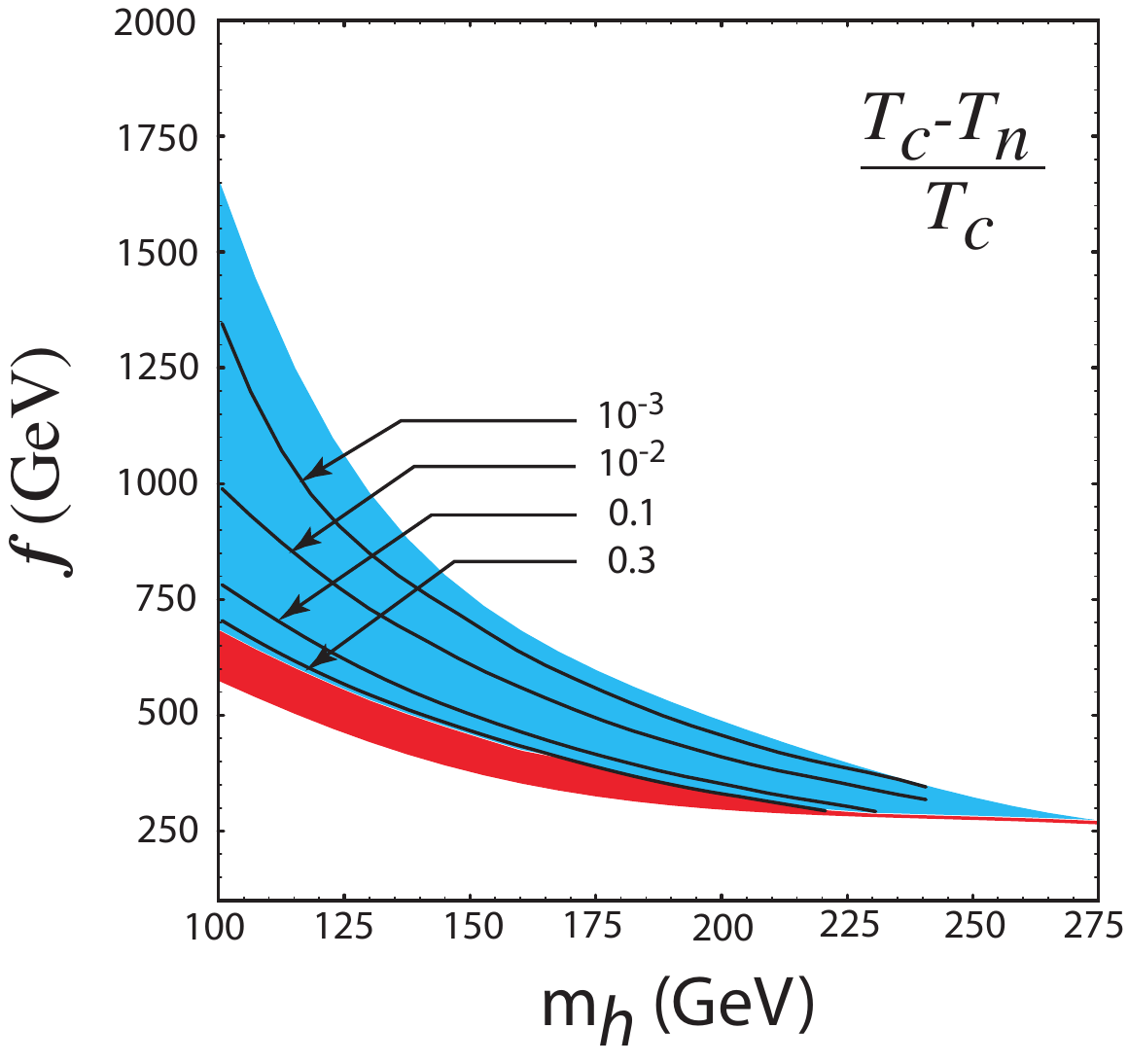}	
		\end{tabular}
	\end{center}
	\caption{The left panel of this figure shows contours of the nucleation temperature $T_n$ in the allowed region for an EW symmetry-breaking first order phase transition ($f$ is the decay constant of the strong sector the Higgs emerges from, and $m_h$ is the physical Higgs mass). Below the red lower bound the EW symmetry remains intact in the vacuum while above the blue upper one the phase transition is second order or not even occurs. Within the red band, the universe is trapped in a metastable vacuum since no expanding bubble is nucleated and the transition never proceeds. The contours are from left to right for $T_n=\{50,100,150\}\gev$. The right panel of this figure shows contours of the relative deviation of the nucleation temperature from the critical one: $\epsilon_T=(T_c-T_n)/T_c$. This measures the degree to which the phase transition is delayed by the overcooling effect. The contours are, from above, for $\epsilon_T=\{10^{-3},10^{-2},0.1,0.3\}$.}
	\label{fig:Tn}
\end{figure}

The contours of constant nucleation temperature are reported in the left panel of Fig.~\ref{fig:Tn}. We point out that there exists a region (painted red in Fig.~\ref{fig:Tn}) with low $f$ and $m_h\lesssim 225\gev$ such that the criterion eq.~(\ref{eqn:Nucleation}) is not satisfied, meaning that the expansion of the universe does not permit the bubbles to percolate. Thus the nucleation never starts and the universe remains trapped in a symmetric vacuum. In addition, the right panel of Fig.~\ref{fig:Tn} helps one to realize further the numerical significance of the supercooling effect by plotting the deviation of the nucleation temperature $T_n$ from $T_c$. We see that, for large values of $f$, the deviation is not significant since the potential barrier disappears at a temperature not much less than the critical one. On the another hand, as soon as one lowers $f$, the barrier persists to lower and lower temperatures, making the supercooling delay of the phase transition important. Thus the knowledge of the nucleation temperature becomes necessary to clearly understand the dynamics of the phase transition in this region.

%%%%%%%%%%%%%%%%%%%%%%%%%%%%%%%%%%%%%%%%%%%%%%%
\subsubsection{Saving the baryon-asymmetry from wash-out}

Understanding the dynamics of the phase transition is a worthy endeavor on its own; however, one of the key reasons for understanding the nature of the EW phase transition is to determine if a baryon asymmetry can be produced and survive the process.  Calculations in the previous sections enable us to 
refine some of the results of \cite{Grojean:2004xa}, where the possibility of a strong first order phase transition was first demonstrated. 

\begin{figure}[t]
\begin{center}
\begin{tabular}{cc}
\includegraphics[width=0.48\hsize]{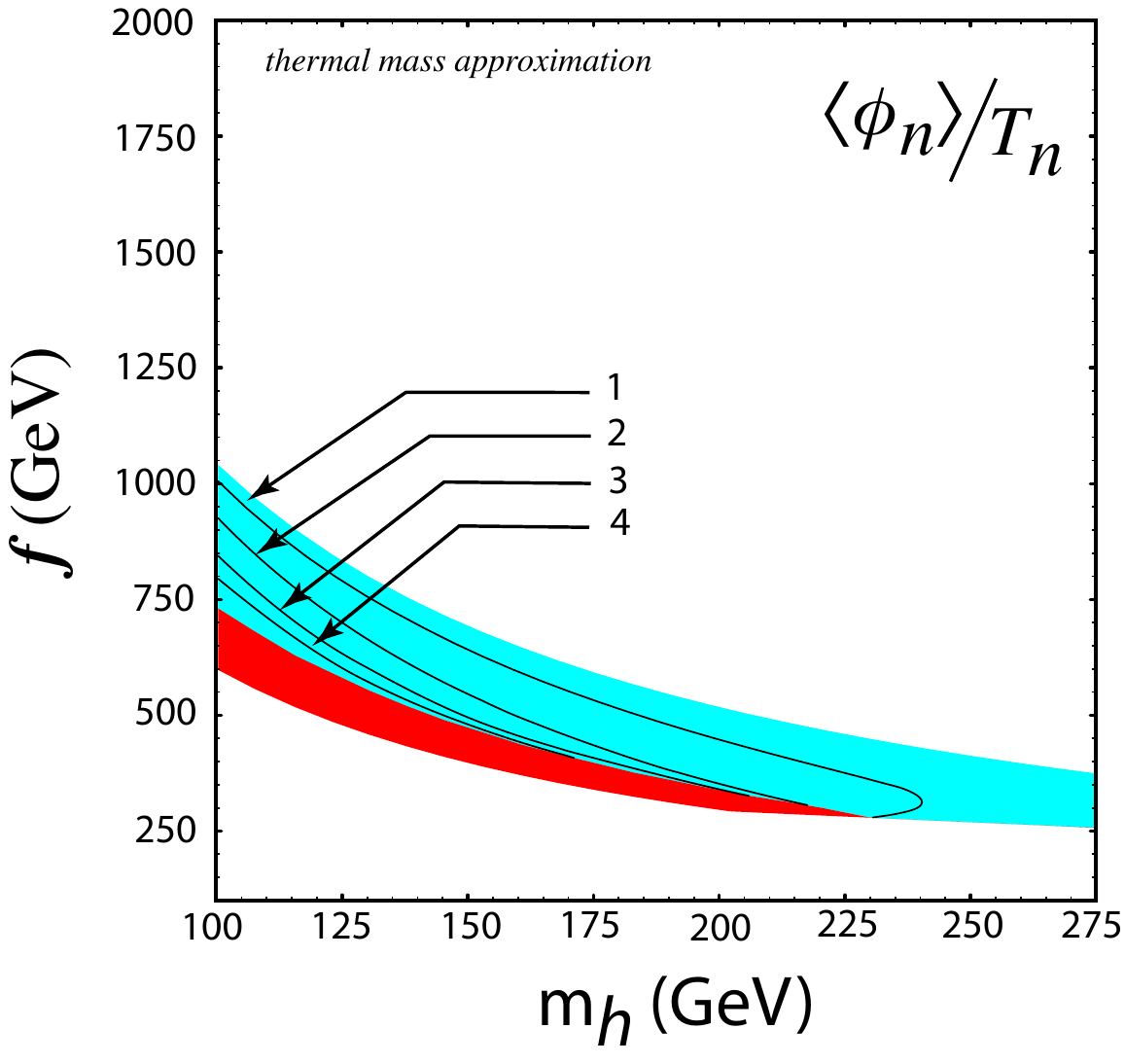}&\includegraphics[width=0.48\hsize]{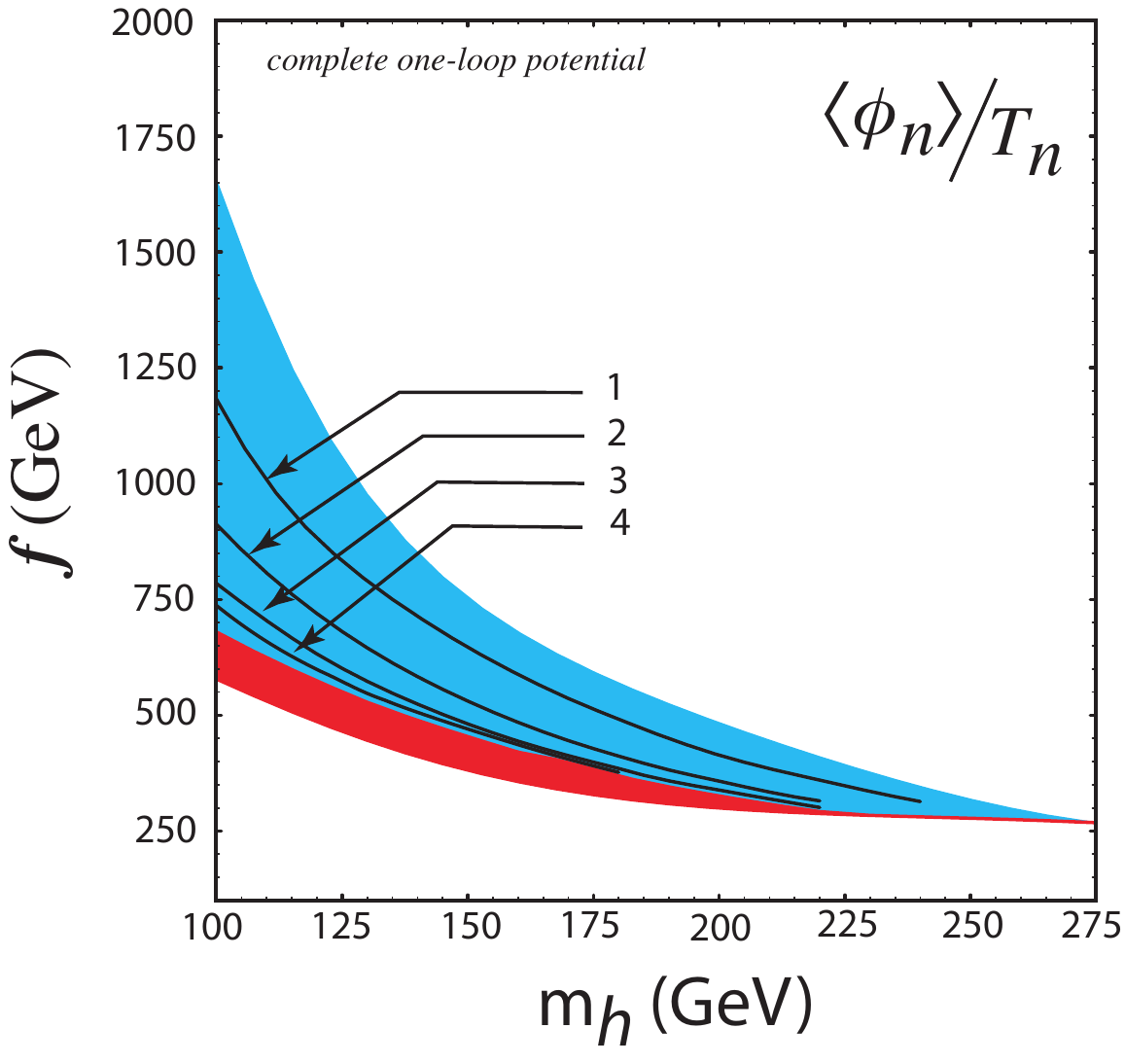}
\end{tabular}
\end{center}
\caption{Plot of the ratio $\xi_n=\langle\phi(T_n)\rangle/T_n$ characterizing the strength of the phase transition using the thermal mass approximation of \cite{Grojean:2004xa} (left) and the complete one-loop potential (right). The contours are for $\xi_n=\{1,2,3,4\}$ from top to bottom. $f$ is the decay constant of the strong sector the Higgs emerges from, and $m_h$ is the physical Higgs mass.}
\label{fig:PhinTn}
\end{figure}

So far we have computed the crucial ratio $\langle\phi(T)\rangle/T$ at the nucleation temperature in the cases where only the thermal masses are included and where the complete one-loop potential is used. This allows us to compare the effect on the wash-out criterion of the supercooling of the phase transition and the usefulness of the one-loop potential. The contour plots of Fig.~(\ref{fig:PhinTn})  show the common fact that the lower the value of $f$, the stronger the phase transition for a fixed Higgs mass. The qualitative result of considering the temperature delay from $T_c$ to $T=T_n$ is that for a given point in the parameters plane, the phase transition is generically stronger at $T_n$. Indeed not only is the nucleation temperature potentially much smaller than $T_c$, but also the value of the Higgs VEV grows as the universe cools down. 

Another important result for the baryon-asymmetry of the universe, is that it can be saved from the wash-out through sphaleron processes, namely $\langle\phi(T)\rangle/T>1$, for a not-so-small value of $f$. Indeed, in order to allow baryogenesis during the EWPT in the approximation of \cite{Grojean:2004xa} some fine-tuning might be required in some approaches without any particular dynamics to make the suppression scale of the dimension six operator in the Higgs sector relatively smaller than the TeV scale required in the gauge sector to pass EW precision measurements. But the full one-loop potential tells us that for values of the Higgs mass above the current experimental bound $f$ can be larger -- as large as $1.2$ TeV -- and the baryon-asymmetry can still freeze out. 
%
%\begin{figure}[!h!t]
%

%%%%%%%%%%%%%%%%%%%%%%%%%%%%%%%%%%%%%%%%%%%

\subsection{Gravitational Waves}

As a bubble expands a part of the latent heat released accelerates the bubble wall and introduces turbulent motions in the hot plasma. After bubbles collide, spherical symmetry is broken. This enables gravitational radiation to be emitted. The turbulence of the plasma after bubble collisions is another important source of gravitational radiation (see~\cite{Buonanno:2007yg} for an introduction to the physics of gravity waves).
In the following, we characterize the spectrum of gravitational radiation that one can expect from the first order phase transition we have detailed in this article.  We compare these results with the sensitivities of current gravity wave detectors, and of proposed gravity wave detectors of the future.

\subsubsection{Characterizing the spectrum}

\begin{figure}[t]
	\begin{center}
		\begin{tabular}{cc}
	\includegraphics[width=0.48\hsize]{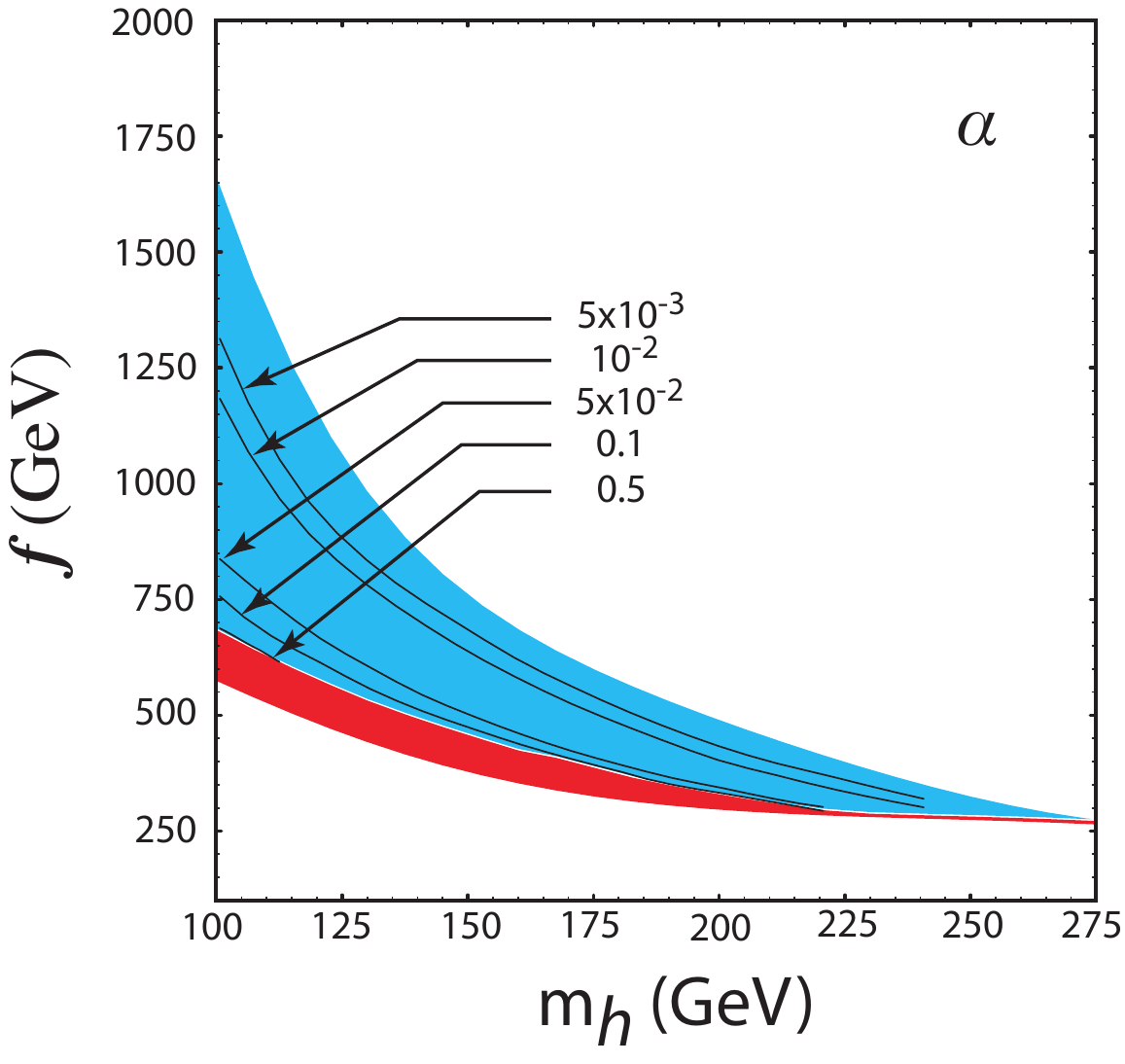}&\includegraphics[width=0.48\hsize]{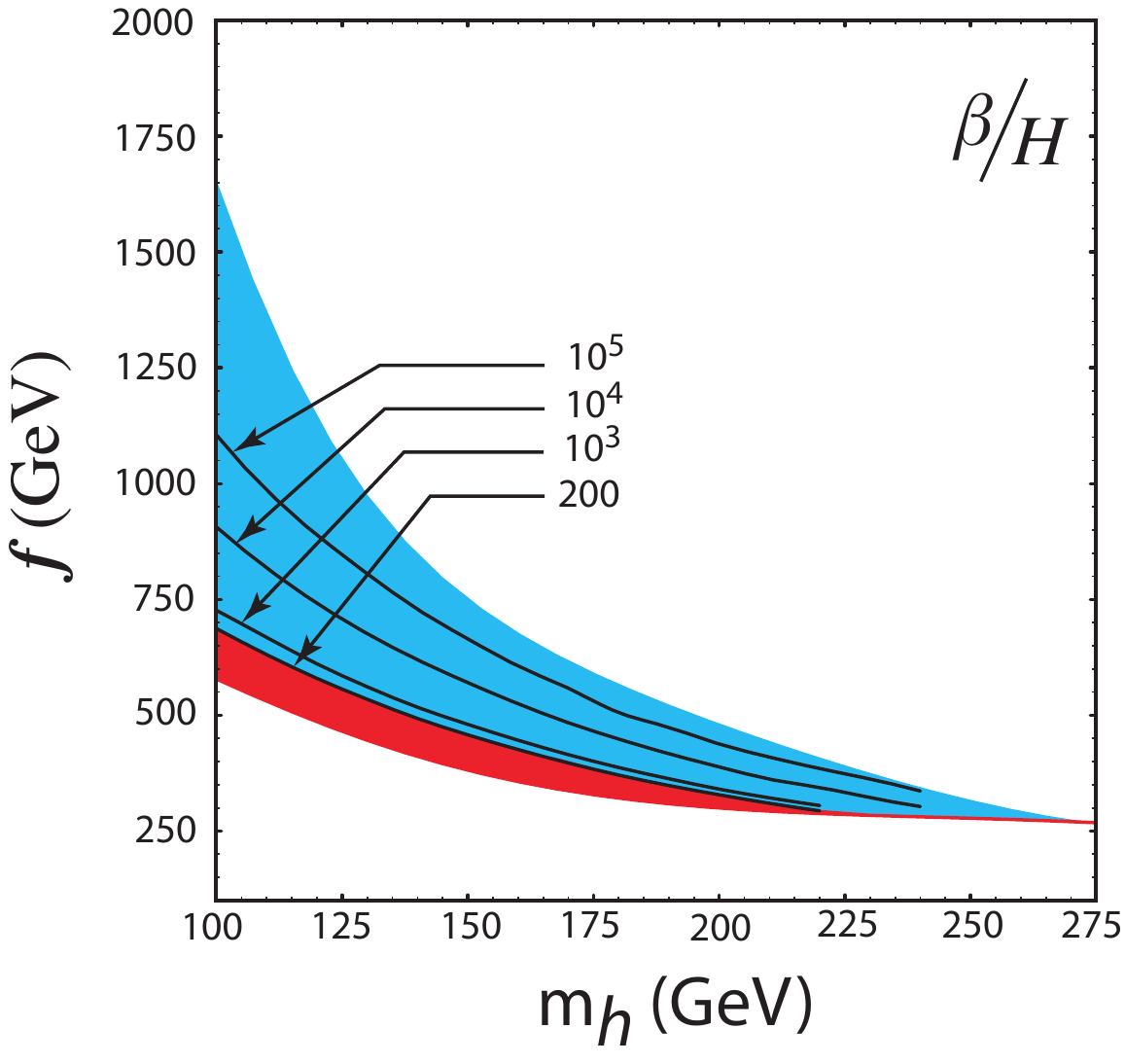}
		\end{tabular}
	\end{center}
	\caption{The panel on the left contains contours of the latent heat $\alpha=\{5.10^{-3},10^{-2},5.10^{-2},0.1,0.5\}$ from top to bottom. The panel on the right draws contours of the parameter, $\beta/H_n$, measuring the duration of the phase transition. From above one has  $\beta/H_n=\{10^5,10^4,10^3,200\}$. $f$ is the decay constant of the strong sector the Higgs emerges from, and $m_h$ is the physical Higgs mass.}
	\label{fig:alpha&beta}
\end{figure}

Previous studies~\cite{Kosowsky,Nicolis,NewGWcomputation} of the gravity wave spectrum culminate in 
showing that it can be fully characterized by the knowledge of only two 
parameters derived ultimately from the 
effective potential\footnote{This conclusion is valid under the assumption of detonation. However, in practice the bubble expand in a thermal bath and not in the vacuum and friction effects taking place in the plasma slow down the bubble velocity. Therefore, it might be important to consider the deflagration regime as in Ref.~\cite{CDS}. When the phase transition is weakly first order, we obtained under the approximations of~\cite{Moore:2000wx} a wall velocity lower than the speed of sound. However, in the interesting region where the phase transition gets stronger, we approach the detonation regime and the approximations of~\cite{Moore:2000wx} have to be refined to accurately compute the wall velocity.
}. The first one is the rate of time-variation of the nucleation rate, named $\beta$. Its inverse gives the duration of the phase transition, therefore defining the characteristic frequency of the spectrum. The second important parameter, $\alpha$, measures the ratio of the latent heat to the energy density of the dominant kind, which is radiation at the epoch considered: $\alpha\equiv\epsilon/\rho_{\rm rad}$. They are both numerically computed from the effective action $S_3/T$ at the nucleation temperature as follows. The time-dependence of the rate of nucleation is mainly concentrated in the effective action and $\beta$ is defined by $\beta\equiv-dS_E/dt\big|_{t_n}$. Using the adiabaticity of the universe one obtain the following dimensionless parameter:
\begin{equation}
\frac{\beta}{H_n}=T_n\frac{d}{dT}\left(\frac{S_3}{T}\right)\Big|_{T_n},
\end{equation}
where $H_n$ is the expansion rate when nucleation starts. The latent energy is the sum of the amount of energy $\Delta V$ seperating the metastable vacuum to the stable one and the entropy variation $\Delta S$ between these two phases. Hence one has:
\begin{equation}
\epsilon=-\Delta V-T\Delta S=\left[-\Delta V+T\frac{\partial V}{\partial T}\right]\Big|_{T_n}.
\end{equation}
The left and right panels of Fig.~\ref{fig:alpha&beta} show contours of constant $\alpha$ and $\beta/H_n$, respectively, at the time of nucleation. 
% 
%\begin{figure}[!h!t!b]

\subsubsection{Observability at interferometry experiments}

Future interferometry experiments could offer us a way to observe the EWPT. A detailed analysis of the potential to directly see gravitational waves from the first-order phase transition
can be compared with the sensitivity expected from 
 the correlated third generation LIGO detector on earth and the LISA and BBO detectors in space. A general analysis that we utilize has been presented in~\cite{Grojean:2006bp}, where both bubble collisions and turbulent motions were considered. 
Qualitatively, gravity-wave detectors will give us a better chance to observe the phase transition today if the latent heat energy released is large and the emission lasts a long time. This can be understood easily by recalling that the power spectrum is given by the square of the quadrupole moment of the source which in turns scales as the kinetic energy over the time of emission \cite{Delaunay:2006ws}. In other words, typically $\alpha$ has to be $\mathcal{O}(1)$ and $\beta/H$ as small as $\mathcal{O}(100)$ to get a sufficiently high energy density $\Omega h^2\gtrsim 10^{-10}$.

Relying on our effective (nonrenormalizable) potential approach, we find  that generically the dynamics of the first order EWPT beyond the SM
generate too weak gravity waves to observe
 except for a tiny region of the parameter space. Namely, by looking closely at Figs.~\ref{fig:alpha&beta}  one can see that for a Higgs mass slightly above the LEP2 bound, $m_h\gtrsim 115$ GeV, and a relatively low scale, $f\sim 650$ GeV, we get at best $\alpha\sim 0.5$ and $\beta/H\sim 100$. The corresponding nucleation temperature in this region is about $50$ GeV, according to Fig~\ref{fig:Tn}. For such a temperature scale, only LISA and BBO will be sensitive to the emitted spectrum of gravity waves, according to the results presented in Figs.~3 and 4 of~\cite{Grojean:2006bp}. Its detectability is probably beyond the capability of LISA. This result is in qualitative agreement 
 with the results of~\cite{Huber:2007vv}.
 Indeed LISA requires at least values of $\alpha>0.6$ for $\beta/H\sim 100$ in order to see the characteristic peak from turbulence while the collision peak starts to be probed for $\alpha>0.8$. On the other hand, BBO should be able to observe both peaks if $\alpha$ is around $0.3$ (keeping $\beta/H\sim 100$). 
 
% \begin{figure}[!h!t!b]
\begin{figure}[t]
	\begin{center}
	\includegraphics[scale=0.8]{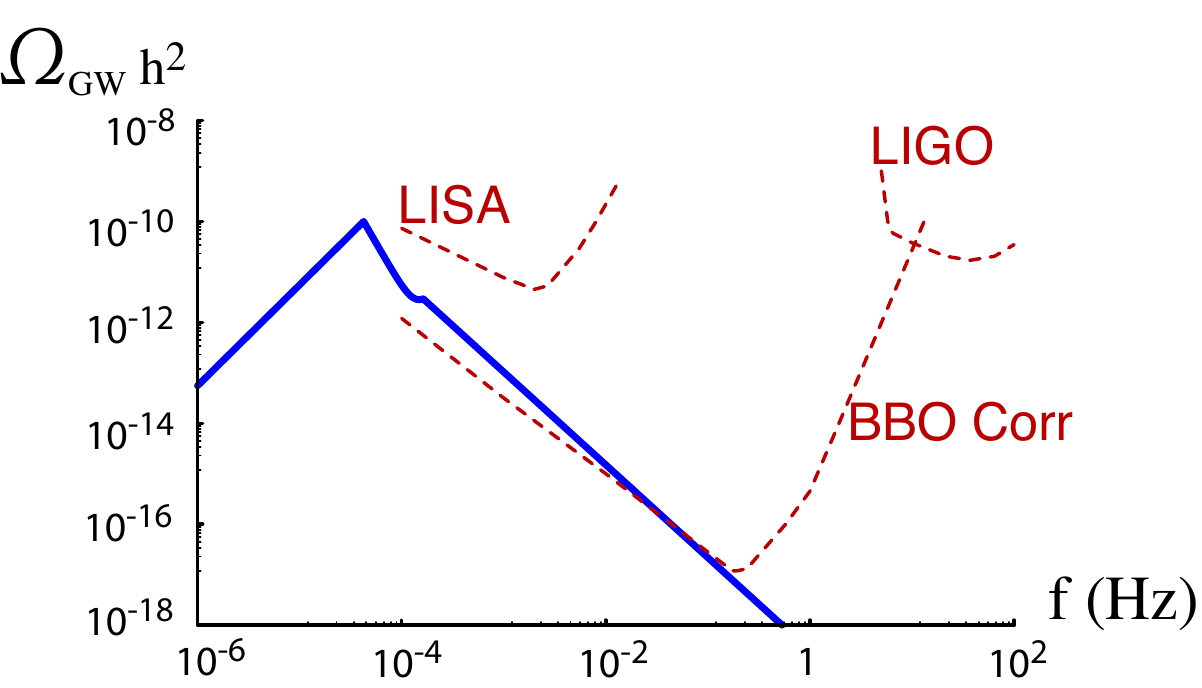}
	\end{center}
	\caption{Example of gravity wave spectrum produced during the EW phase transition both by turbulence (left peak) and collision effects (right peak slightly emerging from the tail of the turbulence spectrum). This plot is for $m_h=115$~GeV and $f\simeq 600$~GeV where $\alpha=0.51$, $\beta/H= 89$ and $T_n=39$~GeV. Note that suitable values of $\alpha,\ \beta/H$ to get a strong signal always imply a small nucleation temperature ($<100$ GeV) due to important overcooling effects that drag the peak below the lower bound of the space-based detectors frequency band ($\simeq 10^{-4}$ Hz), making the gravity waves delicate to observe.}
	\label{fig:GWspectrum}
\end{figure}

Thus it seems that one will have to wait until the launching of the second generation of space-based interferometers to really study the EWPT through gravity wave detectors within this framework.
Moreover this would be possible only in the maximizing case where the Higgs mass is close to its current experimental bound and the composite scale of the Higgs is relatively low.

%%%%%%%%%%%%%%%%%%%%%%%%%%%%%%%%%%%%%%%%%
\section{Conclusions}

In this article we have reported on a complete computation of the one-loop finite temperature effective potential in models where the Higgs boson is composite and emerges as a light pseudo-Goldstone boson of a strongly interacting sector (our analysis could also be relevant for studying the dynamics of electroweak symmetry breaking in Little Higgs theories). These models are characterized by higher dimensional operators in the Higgs sector suppressed by the strong decay constant, $f$, a scale parametrically smaller than the cutoff of the strong sector. Interestingly, by following the details of the phase transition dynamics, the parameter space of a strong first-order phase transition has actually grown for large value of $f$, and shrunk for small value of $f$ cutoff, compared to the tree-level result found in~\cite{Grojean:2004xa}.  It has grown at the higher end by going beyond the high temperature approximation.  The parameter space has shrunk on the lower end, since we found that bubbles cannot be nucleated well enough there to overcome the effects of an expanding universe. We encountered some subtleties along the way, including infrared singularities and imaginary components to the potential, that were resolved.

It was also necessary to compute the details of the phase transition dynamics in order to investigate the possibility of detecting gravitational radiation from the first order phase transition occuring in the early universe. After bubbles are nucleated, their collisions and subsequent turbulence in the plasma give rise to gravity waves.  In the assumption of a detonation regime, the effects depend on only two parameters, the latent heat $\alpha$ and the duration of the phase transition $\beta^{-1}$, both of which can be determined by solving the bounce equation, and analyzing the full one-loop finite temperature effective potential at the scale of the nucleation temperature.  Although LIGO and LISA are likely not sensitive to these effects, we found that BBO, a planned second generation experiment of space-based interferometers, could be sensitive to the gravity waves produced during this phase transition.

%%%%%%%%%%%%%%%%%%%%%%%%%%%%%%%%%%%%%%%%%
\section*{Acknowledgments}
We thank J.R.~Espinosa, S.~Martin, M.~Perelstein, M.~Serone and G.~Servant for helpful conversations. We also want to thank T.~Hahn for his help with the LoopTools software. This work is supported in part by the Department of Energy and the Michigan Center for
Theoretical Physics (MCTP), by  the RTN European Program MRTN-CT-2004-503369, by the EU FP6 Marie Curie RTN ``UniverseNet" (MRTN-CT-2006-035863) and by the CNRS/USA exchange grant 3503.

\appendix

%%%%%%%%%%%%%%%%%%%%%%%%%%%%%%%%%%%%%%%%%
\section{On-Shell Renormalization of the $T=0$ Potential}
\label{sec:T=0corrections}

The on-shell scheme identifies lagrangian parameters as physical parameter (i.e., observables). 
It is the scheme employed by~\cite{Anderson:1991zb}, although we augment that discussion by describing a self-consistent approach with higher order operators, and describe the details of how IR divergences from massless Goldstone bosons cancel. 

Renormalizing our theory in the on-shell scheme is most convenient when we begin by writing the full potential in the following form:
\beq
V_{tree}(\phi)=\frac{\lambda}{4}\left( \phi^2-v_0^2\right)^2+\frac{\kappa}{8}\left( \phi^2-v_0^2\right)^3 +\Delta V_1^0(\phi)+\frac{\delta c_2}{2}\phi^2+\frac{\delta c_4}{4}\phi^4
+\frac{\delta c_6}{6}\phi^6
\eeq
where 
\begin{eqnarray}
\Delta V_1(\phi,T=0)\equiv \Delta V_{1}^0(\phi)&=&\sum_{i=h,\chi,W,Z,t}\frac{n_i}{2}\int \frac{d^4 k_E}{(2\pi)^4}\log\left[k_E^2+m^2_{i}(\phi)\right]\label{eqn:zeroTcorrection}\\
&=&\sum_{i=h,\chi,W,Z,t}n_i\frac{m_{i}^4(\phi)}{64\pi^2}\left[\log\frac{m^2_{i}(\phi)}{\mu^2}-C_i-C_{UV}\right]\label{eqn:dimregpot}
\end{eqnarray}
which has been regularized in $4-\epsilon$ dimensions, $C_i=5/6$ ($3/2$) for gauge bosons (scalars and fermions) and $C_{UV}\equiv\frac{2}{\epsilon}-\gamma_E+\log 4\pi+{\cal O}(\epsilon)$.
In this parametrization of the tree-potential, the scalar $\phi$-dependent masses are: $m^2_h(\phi)=\lambda(3\phi^2-v_0^2)+3\kappa(5\phi^4-6v_0^2\phi^2+v_0^4)/4$ and $m^2_\chi(\phi)=\lambda(\phi^2-v_0^2)+3\kappa(\phi^2-v_0^2)^2/4$.
The on-shell scheme imposes that $v_0$ is the vacuum expectation of the Higgs field, $\lambda \equiv \frac{m^2_h}{2v_0^2}$, and $\kappa\equiv 1/f^2$.  The precise meaning of $f^2$ is defined below. 

The counter terms, $\delta c_i$, are determined by the renormalization conditions:
\begin{eqnarray}
\frac{dV_{eff}(\phi,T=0)}{d\phi}\Big|_{\phi=v_0}&=&0,\label{eqn:RC1a}\\
\frac{d^2 V_{eff}(\phi,T=0)}{d\phi^2}\Big|_{\phi=v_0}&=&m^2_h-\Delta\Sigma,\label{eqn:RC1b} \\
\frac{d^3 V_{eff}(\phi,T=0)}{d\phi^3}\Big|_{\phi=v_0}&=& \xi_{phys}-\Delta\Gamma\label{eqn:RC1c}
\end{eqnarray}
where $\Delta\Sigma=\Sigma(m_h)-\Sigma(0)$ and $\Delta\Gamma=\Gamma(m_h)-\Gamma(0)$
are needed to take us from the IR-sensitive and unphysical $p=0$ limit of the effective potential to $p^2=m_h^2$, where physical observables $m_h$ and the tri-Higgs coupling $\xi_{phys}$ are defined.
Detailed computations demonstrating the cancelation of the IR divergences in this scheme are presented in Appendix~\ref{app:onshell}.

We wish to have a more direct physical parameter that  parametrizes deviations from the SM, and so we redefine 
\beq\label{eqn:xidef}
\xi_{phys}\equiv \xi^{SM}_{phys}+\frac{6v_0^3}{f^2}
\eeq
which constitutes the definition of the physical observable $f$.  Recall that the tri-Higgs coupling in the SM is fixed with knowledge of $m_h$, and thus $\xi^{SM}_{phys}$ is determined completely by $m_h$ and the other parameters of the SM:
\beq
\xi^{SM}_{phys}= \frac{3m_h^2}{v_0}+\sum_i \frac{n_i}{32\pi^2}\frac{[m_i^2(v_0)']^3}{m_i^2(v_0)}.
\eeq
Since $m^2(v_0)$ depends on $1/f^2$, this expression is technically equal to the SM one only in the limit of $f^2 \to \infty$, which is all that we need for the analysis to be self-consistent.

We are now able to invert the renormalization conditions and compute the counter terms, which depend on the various derivatives of $V_1^0(\phi)$, $\Delta\Sigma$, and $\Delta\Gamma$.  Upon expanding the result, one can express the renormalized full one-loop potential as
\bea
V_{eff}(\phi)& = &\frac{m_h^2}{8v_0^2}(\phi^2-v_0^2)^2+\frac{1}{8f^2}(\phi^2-v_0^2)^3 \\
& & + \sum_{i=h,\chi,W,Z,t}\frac{n_i}{64\pi^2}\left[m^4_i(\phi)\left(\log\frac{m^2_i(\phi)}{m^2_i(v_0)}-\frac{3}{2}\right)+2 m^2_i(v_0) m^2_i(\phi)\right] \nonumber\\
& & + \frac{1}{16}\left( 7\Delta\Sigma-v_0\Delta \Gamma\right) \phi^2
+\frac{1}{16v_0^2}\left( -5\Delta\Sigma+v_0\Delta\Gamma\right) \phi^4
+ \frac{1}{48v_0^4}\left( 3\Delta\Sigma -v_0\Delta\Gamma\right)\phi^6 \nonumber
\eea
where all the $f$-dependence of the loop-order contribution to the potential is contained in the field-dependent masses, making the continuity of the decoupling limit explicit.

%%%%%%%%%%%%%%%%%%%%%%%%%%%%%%%%%%%%%%%%%%%%%%%%%%%
\section{Cancelation of Goldstone Boson IR Divergences}\label{app:onshell}
In this Appendix, we gather the detailed computations for the results mentioned in Appendix~\ref{sec:T=0corrections} about smoothing the Goldstone IR singularity in the one-loop potential at zero temperature. First, we shall briefly recall how one moves from zero-momentum to on-shell scheme in the SM, as a warm-up for the dimension-six operator discussion that will come afterwards.
\subsection{Review of the SM case}
In the SM the loop-integral of (\ref{eqn:zeroTcorrection}) can be renormalized by imposing the two conditions
\begin{eqnarray}
\frac{dV_{eff}(\phi,T=0)}{d\phi}\Big|_{\phi=v_0}&=&0,\\
\frac{d^2 V_{eff}(\phi,T=0)}{d\phi^2}\Big|_{\phi=v_0}&=&m^2_{h,0}\label{eqn:RC2off},
\end{eqnarray}
which leads to the traditional form of the effective potential
\bea
V^{(SM)}_{eff}(\phi)& = &\frac{m_{h,0}^2}{8v_0^2}(\phi^2-v_0^2)^2\\
& & + \sum_{i}\frac{n_i}{64\pi^2}\left[m^{4}_i(\phi)\left(\log\frac{m^{2}_i(\phi)}{m^{2}_i(v_0)}-\frac{3}{2}\right)+2 m^{2}_i(v_0) m^{2}_i(\phi)\right] \nonumber
\eea
where the scalar masses must be evaluated in the decoupling limit of the dimension-six operator ($f\rightarrow\infty$).
As we will review, $m_{h,0}$ is an off-shell Higgs mass defined at $p=0$. This is fine to use  as long as no massless particle couples to the Higgs field \cite{Anderson:1991zb}. If such particles like the Goldstone bosons are to be taken into account, one must move away from zero-momentum to avoid the pole-mass at $p=0$ that makes both $m_{h,0}$ and the one-loop potential IR divergent.

In order to see how this can be done, we recall that near the symmetry breaking minimum ($\phi=v_0$) the renormalized effective potential can always be expanded in terms of 1PI-Green functions evaluated at vanishing external momentum as follow:
\begin{equation}\label{eqn:appVeffexpansion}
V_{eff}(\phi,T=0)=-\sum_{n=0}^\infty \frac{(\phi-v_0)^n}{n!}\mathcal{G}^{(n)}({p_i^2=0}),
\end{equation}
where $\mathcal{G}^{(n)}({p_i^2})$ are the $n$-legs renormalized 1PI Green functions for the physical Higgs scalar evaluated about the true vacuum (i.e., in the shifted theory). This expansion directly follows from the fact that the effective action may be intepreted as a generating functional of these 1PI Green functions. Hence the second derivative of the effective potential at $v_0$ is simply the renormalized two-point function at zero-momentum:
\begin{equation}\label{eqn:appd2Veff}
\frac{d^2 V_{eff}(\phi,T=0)}{d\phi^2}\Big|_{\phi=v_0}=-\mathcal{G}^{(2)}(p^2=0).
\end{equation}
Given that the two-point function (the inverse propagator) of the Higgs is
\begin{equation}
\mathcal{G}^{(2)}(p^2)=p^2-\left(m^2_{h,R}+\Sigma(p^2)\right)\label{eqn:app2ptfunction},
\end{equation}
where $m^2_{h,R}$ and $\Sigma(p^2)$ are the renormalized Higgs mass and one-loop Higgs self-energy, we see that imposing the renormalization condition (\ref{eqn:RC2off}) leads to
\begin{equation}
m_{h,0}^2=m_{h,R}^2+\Sigma(p^2=0).
\end{equation}
justifying that $m_{h,0}$ is to be understood as the zero-momentum Higgs mass. In order to circumvent the IR divergences a natural choice would be to express the right-hand side of (\ref{eqn:RC2off}) in terms of physical parameters. The physical Higgs mass ($m_h$) is defined as the pole of the one-loop resummed propagator ($\mathcal{G}^{(2)}(p^2=m_{h}^2)=0$) and is given by solving the self-consistent equation:
\begin{equation}\label{eqn:appmhphys}
m_{h}^2=m_{h,R}^2+\Sigma(p^2=m_{h}^2).
\end{equation}
This allows us to rewrite (\ref{eqn:appd2Veff}) as
\begin{eqnarray}
\frac{d^2 V_{eff}(\phi,T=0)}{d\phi^2}\Big|_{\phi=v_0}&=&m_{h,R}^2+\Sigma(p^2=0),\nonumber\\
&=&m_{h}^2-\Delta\Sigma
\end{eqnarray}
with $\Delta\Sigma\equiv\Sigma(p^2=m_{h}^2)-\Sigma(p^2=0)$.

\begin{figure}[t]
	\centering
	\includegraphics[scale=0.5]{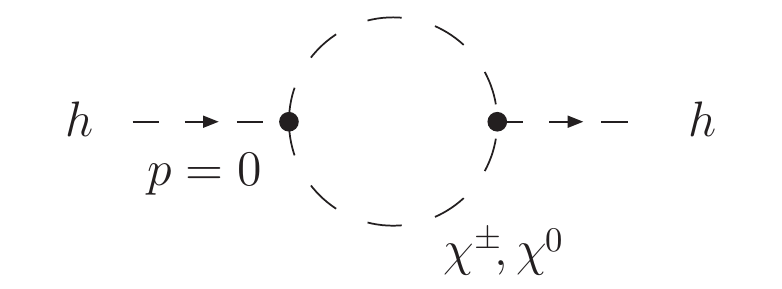}
	\caption{IR divergent diagram contributing to $\Sigma(p^2=0)$.}
	\label{fig:2ptIRdiv}
\end{figure}

When $\Delta\Sigma$ is absent the UV-finite one-loop correction at zero temperature for the SM 
has an IR divergent piece coming from the Goldstone contribution:
\begin{equation}\label{eqn:V1SMIRdiv}
\Delta V_{1}^{0\,(SM)}(\phi)^{(IR\,div)}=-\frac{n_\chi}{64\pi^2}m_\chi^{4\,(SM)}(\phi)\log m^2_\chi\underset{m^2_\chi\to 0}{\longrightarrow} \infty
\end{equation}
where $m^2_\chi\equiv m^{2\,(SM)}_\chi(v_0)$ will be kept non-zero as a regulator in what follows. Moving to the on-shell renormalization scheme (i.e., replacing (\ref{eqn:RC2off}) by (\ref{eqn:RC1b})) results in the addition of the following term to $V_{eff}^{(SM)}$:
\begin{equation}\label{eqn:extraV1RSM}
\delta V_{1,(SM)}(\phi)=-\frac{\Delta\Sigma^{(SM)}}{8 v_0^2}(\phi^2-v_0^2)^2
\end{equation}
Now $\Sigma^{(SM)}(p^2=0)$ receives an IR singularity from the diagram depicted in Fig.~\ref{fig:2ptIRdiv} which can be easily calculated to give:
\begin{equation}\label{eqn:Amp2ptIRdiv}
\Sigma_{(IR\,div)}^{(SM)}(p^2=0)=\frac{n_\chi}{32\pi^2}\frac{m_h^4}{v_0^2}\log m^2_\chi.
\end{equation}
Combining (\ref{eqn:Amp2ptIRdiv}) with (\ref{eqn:extraV1RSM}), one gets the following IR divergent contribution to the potential (up to an irrelevant $\phi$-independent term):
\begin{equation}
\delta V_{1,(SM)}(\phi)^{(IR\,div)}=\frac{n_\chi}{64\pi^2}m^{4\,(SM)}_\chi(\phi)\log m^2_\chi,
\end{equation}
which exactly cancels out (\ref{eqn:V1SMIRdiv}).
\subsection{Generalization to the non-renormalizable potential}
The presence of the dimension six interaction at tree-level forces us to set one more derivative of the potential 
to an extra measurable quantity. Focusing only on the decoupling limit this can be accomplished by
\bea
\frac{d^3 V_{eff}(\phi,T=0)}{d\phi^3}\Big|_{\phi=v_0}&=&\xi_0\label{eqn:RC3off},
\eea
where
\bea\label{eqn:xidef2}
\xi_0\equiv \frac{3m^2_{h,0}}{v_0}+\frac{6v_0^3}{f_0^2}+\sum_i\frac{n_i}{32\pi^2}\frac{[m^2_i(v_0)']^3}{m^2_i(v_0)} .
\eea

As can be easily checked, defining only the Higgs mass on-shell does not smooth out entirely the bad IR behavior of the one-loop potential in the non-renormalizable case. In fact $f$, being another parameter to be fixed at the quantum level, needs also to be renormalized away from zero-momentum to avoid the Goldstone pole, which is done by defining the renormalized three-point function of the Higgs boson on-shell. As for the two-point function, from (\ref{eqn:appVeffexpansion}) one gets
\begin{equation}
\frac{d^{3} V_{eff}(\phi,T=0)}{d\phi^{3}}\Big|_{\phi=v_0}=-\mathcal{G}^{(3)}(p_i^2=0)
\end{equation}
where $p_i$ denotes the external momenta of the three-point function. $\mathcal{G}^{(3)}$ can be split into a tree-level coupling and a one-loop correction as
\begin{equation}
\mathcal{G}^{(3)}(p_i^2)=-g_{3}-\Gamma_3(p_i^2)
\end{equation}
where $g_3$ is the renormalized cubic self-couplings of the Higgs at tree-level. Similarly to the Higgs mass, we see that imposing (\ref{eqn:RC3off}) implies working with a parameter $\xi_0$, or rather $f_0$ through (\ref{eqn:xidef2}), defined at zero-momentum which leads again to IR divergent behavior. We propose defining an on-shell cubic coupling at one-loop by\footnote{Other physical definitions of the cubic coupling are possible, so long as they move away from zero-momentum to solve the IR issue.}
\begin{equation}
\xi_{phys}\equiv -\mathcal{G}^{(3)}(p_i^2=m_{h}^2)=g_3+\Gamma_3(p_i^2=m_{h}^2),
\end{equation}
which translates into an on-shell (physical) definition of $f$ by means of (\ref{eqn:xidef}). Finally by expressing (\ref{eqn:RC3off}) in terms of physical parameters, we get the on-shell renormalization condition of (\ref{eqn:RC1c})
\begin{eqnarray}
\frac{d^{3} V_{eff}(\phi,T=0)}{d\phi^{3}}\Big|_{\phi=v_0}&=&g_3+\Gamma_3(p_i^2=0),\\
&=&\xi_{phys}-\Delta\Gamma_3
\end{eqnarray}
where $\Delta\Gamma\equiv \Gamma_3(p_i^2=m^2_{h})-\Gamma_3(p_i^2=0)$.

\begin{figure}[t]
	\centering
	\includegraphics[scale=0.5]{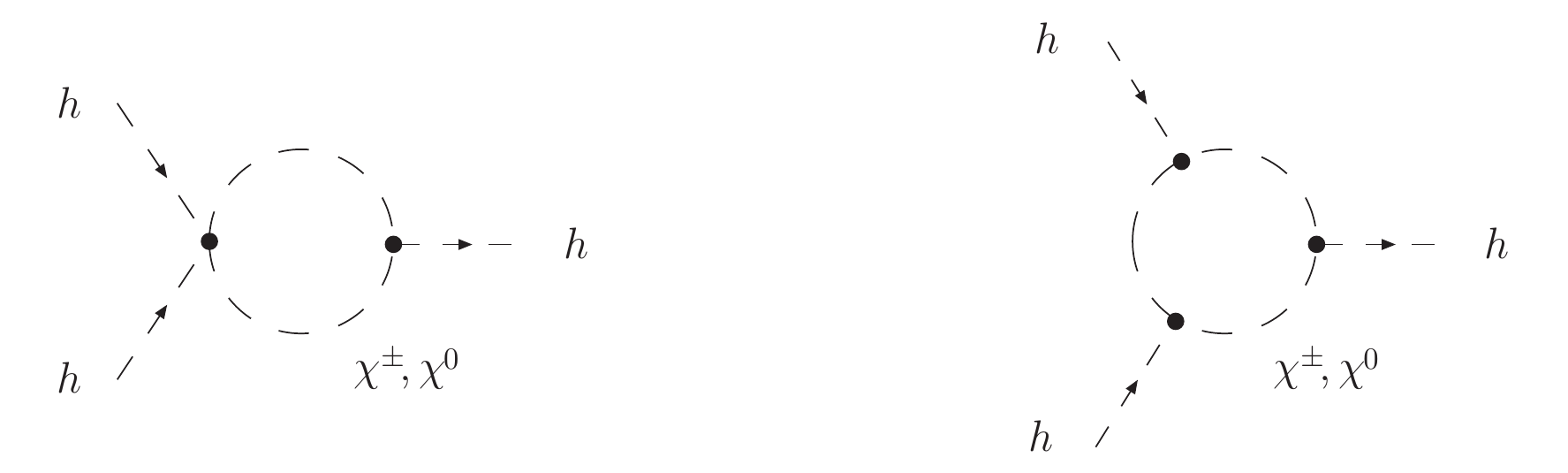}
	\caption{IR divergent diagram contributing to $\Gamma_{3}(p_i^2=0)$. All the momenta are assumed to be zero in the external lines.}
	\label{fig:3ptIRdiv}
\end{figure}

Now by enforcing the three renormalization conditions (\ref{eqn:RC1a}),(\ref{eqn:RC1b}) and (\ref{eqn:RC1c}) to set the counter-terms, we find that the zero-momentum potential
is augmented by
\begin{eqnarray}
\delta V_{1}(\phi)=-\frac{\Delta\Sigma}{8 v_0^2}\left(\phi^2-v_0^2\right)^2+\left[\frac{\Delta\Sigma}{16 v_0^4}-\frac{\Delta\Gamma}{48 v_0^3}\right]\left(\phi^2-v_0^2\right)^3
\end{eqnarray}
We recall that in terms of $m_{h,0}$ and $\xi_0$ the effective potential develops a logarithmic IR singularity of the same form as in the SM but with $f$-dependent masses:
\begin{equation}\label{eqn:appV1OFFIRdiv}
\Delta V_1^{0}(\phi)^{(IR\,div)}=-\frac{n_\chi}{64\pi^2}m_\chi^{4}(\phi)\log m^2_\chi,
\end{equation}
while $\xi_0$ defined in (\ref{eqn:xidef2}) has a power-law divergence. Nonetheless $\Gamma_{3}(p_i^2=0)$ contains IR-divergent parts from the diagrams of Fig.~\ref{fig:3ptIRdiv} which are
\begin{eqnarray}
\Gamma_{3}(p_i^2=0)^{(IR\,div)}&=&\frac{3n_\chi}{32\pi^2}\frac{m_h^4}{v_0^3}\left(1+\frac{6v_0^4}{m_h^2f^2}\right)\log m^2_\chi\\
& &+\frac{n_\chi}{32\pi^2}\frac{m_h^6}{m_\chi^2 v_0^3}.
\end{eqnarray}

Hence the power-law divergence of $\Delta\Gamma$ cancels with the one of (\ref{eqn:xidef2}) making $\xi_{phys}$ a well-defined quantity. 
The remaining logarithmic divergence of the three-point function along with the one from the self-energy, which turns out to be the same as in the SM,
\begin{equation}
\Sigma_{(IR\,div)}(p^2=0)=\frac{n_\chi}{32\pi^2}\frac{m_h^4}{v_0^2}\log m^2_\chi,
\end{equation}
gives after some simple algebra (up to irrelevant constant and $\mathcal{O}(f^{-4})$ terms)
\begin{equation}
\delta V_{1}(\phi)^{(IR\,div)}=\frac{n_\chi}{64\pi^2}m^{4}_\chi(\phi)\log m^2_\chi
\end{equation}
which cancels with eq.~(\ref{eqn:appV1OFFIRdiv}).
 
Finally, one finds that this procedure also leads to a UV and IR finite potential when a different higher derivative is chosen as a third renormalization condition.

%%%%%%%%%%%%%%%%%%%%%%%%%%%%%%%%%%%%%%%%%%%
%%%%%%%%%%%%%%%%%%%%%%%%%%%%%%%%%%%%%%%%%%%%%%%%%%%%%%%%%%%%%%

%%%%%%%%%%%%%%%%%%%%%%%%%%%%%%%%%%%%%%%%%
\section{Review of $T\neq 0$ One-Loop Higgs Potential}\label{app:review_technics}
\subsection{The one-loop potential from the background field method}
The original method proposed by Jackiw in \cite{Jackiw:1974cv} to compute loop corrections to the classical potential is based upon expanding the action about (constant) background values for the various fields appearing in the theory. In our case only the neutral Higgs component has a non-vanishing VEV, and we recall here in a concise way how this method allows one to derive the one-loop correction given in (\ref{eqn:fullV1eff}). At the start, we consider the $T=0$ correction, and discuss the finite $T$ corrections in the next subsection of this Appendix.

As an illustration we focus on a simple self-interacting scalar (real) field theory defined by the following generating functional:
\begin{equation}\label{eqn:appZj}
Z[j]\equiv \int [\mathcal{D}\phi]\mbox{exp}\left[i(S[\varphi]+j\varphi)\right]
\end{equation}
where the notation $\varphi j\equiv\int d^4x\varphi(x)j(x)$ will be assumed throughout this Appendix, and the action is  $S[\varphi]=\int d^4x[(\partial_\mu\varphi)^2/2-V_0(\varphi)]$. Then one shifts the field by a $x$-independent background value ($\varphi(x)=\phi+h(x)$) where $\phi$ is assumed to be a classical field configuration and $h$ represents a quantum fluctuation about it. We will now integrate out this fluctuation to get its effect on the potential up to one-loop order. To do so, one defines the shifted theory (whose dynamical field is now $h$) by expanding the action about its classical value:
\begin{equation}
S[\phi+h]+j(\phi+h)=S[\phi]+j\phi+h\left(\frac{\delta S}{\delta \varphi}\Big|_{\varphi=\phi}+j\right)+\frac{1}{2}h_x\frac{\delta^2 S}{\delta\varphi_x\delta\varphi_y}\Big|_{\varphi=\phi}h_y+\cdots
\end{equation}
where thanks to the equation of motion in the presence of a source the linear term vanishes.  The $\cdots$ stand for higher (than quadratic) orders in $h$ which lead to (at most) two-loop corrections \cite{Jackiw:1974cv}. One also easily obtains after an integration by parts that
\begin{equation}\label{eqn:appD2S}
\frac{\delta^2 S}{\delta\varphi_x\delta\varphi_y}\Big|_{\varphi=\phi}=-(\Box+V_0''(\phi))\delta^4(x-y).
\end{equation}
Plugging this expansion back into (\ref{eqn:appZj}) one obtains
\begin{eqnarray}
Z[j]&\simeq& e^{i(S[\phi]+j\phi)}\times\int[\mathcal{D}h]\exp\left[\frac{i}{2}h_x\frac{\delta^2 S}{\delta\varphi_x\delta\varphi_y}\Big|_{\varphi=\phi}h_y\right]\nonumber\\
&=&e^{i(S[\phi]+j\phi)}\times\mbox{Det}\left(\Box+V_0''(\phi)\right)^{-\frac{1}{2}}.
\end{eqnarray}
We recall that by definition the effective action is the Legendre-transform of the logarithm of $Z[j]$
\begin{equation}
S_{eff}[\phi]\equiv-i\log Z[j]-j\phi,
\end{equation}
which in our case, including the quantum fluctuations at one-loop, takes the form:
\begin{equation}\label{eqn:appSeff}
S_{eff}[\phi]=S[\phi]+\frac{i}{2}Tr\log\left(-\frac{\delta^2 S}{\delta\varphi_x\delta\varphi_y}\Big|_{\varphi=\phi}\right).
\end{equation}
Moreover $S_{eff}$ can always admit a derivative expansion of the form:
\begin{equation}
S_{eff}[\phi]\equiv\int d^4x\left[-V_{eff}(\phi)+A(\phi)(\partial_\mu \phi)^2+\cdots\right],
\end{equation}
which defines precisely what one calls the effective potential. Since $\phi$ is an homogeneous configuration in space-time, this simplifies to:
\begin{equation}
S_{eff}[\phi]=-\mathcal{V} V_{eff}(\phi),
\end{equation}
where $\mathcal{V}$ is the volume of space-time we choose to keep finite for the moment. Besides, this homogeneity preserves the diagonality of  (\ref{eqn:appD2S})  in momentum-space, which allows us to evaluate the trace in (\ref{eqn:appSeff}). This leads to the following expression for the effective potential:
\begin{equation}
V_{eff}(\phi)=V_0(\phi)-\frac{i}{2}\mathcal{V}^{-1}\sum_k\log(-k^2+V_0''(\phi))
\end{equation}
where the sum is over the eigenvalues of the $\Box$ operator in momentum-space. Finally by taking the limit of infinite space-time volume, one gets the well-known result:
\begin{equation}\label{eqn:appV1}
V_{eff}(\phi)=V_0(\phi)-\frac{i}{2}\int\frac{d^4k}{(2\pi)^4}\log(-k^2+V_0''(\phi)).
\end{equation}
The generalization for fields of higher spin that couple to $\phi$ is 
\begin{equation}
V_{eff}(\phi)=V_0(\phi)+i\sum_{i={\rm fields}}\eta\int\frac{d^4k}{(2\pi)^4}\log\mbox{det}\left(-i\widetilde{D}_i(k,\phi)\right),
\end{equation}
where $-i\widetilde{D}(k,\phi)$ is the inverse propagator, $\eta=-1/2\,(1)$ for bosons (fermions) is the power of the functional determinant, and the $\mbox{det}$ denotes an eventual determinant acting on either Lorentz or Dirac indices. 
%%%%%%%%%%%%%
\subsection{Turning on the temperature in the effective potential}
The imaginary time formalism  to go from quantum statistics at zero-temperature to thermal quantum statistics is by compactification of the euclidean time dimension on a circle of radius $R=1/2\pi T$. This correspondence is formally obtained in the path integral formulation of quantum 
mechanics~\cite{LeBellac}. However, it is worthwhile to give a quick intuitive argument. 

We begin with the generating functional for a scalar field in euclidean space-time ($\tau=it$):
\begin{equation}
Z[j]=\int [\mathcal{D}\phi] \exp\left[-\int d^4x_E\left(\frac{1}{2}\partial_\mu\phi\partial^\mu\phi+V_0(\phi)+j\phi\right)\right].
\end{equation}
Now requiring the euclidean time to lie in the interval $-1/2T\leqslant\tau\leqslant 1/2T$, and restricting the field $\phi$ to static configurations, one ends up with
\begin{equation}
Z[j]=\int [\mathcal{D}\phi] \exp\left[-\frac{1}{T}\int d^3x\left(\frac{1}{2}\partial_i\phi\partial^i\phi+V_0(\phi)+j\phi\right)\right].
\end{equation}
For vanishing source the space integral is nothing else but the energy ($E[\phi]$) stored in a (time-independent) field configuration $\phi$, and the generating functional reduces to
\begin{equation}
Z[j=0]=\int [\mathcal{D}\phi]\ e^{-\frac{E[\phi]}{T}}\sim \sum_{S={\rm all\, states}}e^{-E_S/T},
\end{equation}
which is the common partition function of statistical mechanics where $\phi$ describes all possible (static) configurations of a given system in equilibrium with a heat reservoir at temperature $T$.

Therefore the prescription to follow as soon as temperature is switched on is rather simple. It consists of Fourier expanding the fields among its eigen (Matsubara) frequencies $\omega_n$ and discretizing the imaginary time integrals by  the following replacement rule:
\begin{equation}\label{eqn:Trule}
\int\frac{dk_{0,E}}{2\pi}f(k_{0,E})\rightarrow T\sum_{n=-\infty}^{\infty}f(k_{0,E}=\omega_n).
\end{equation}
For instance, applying (\ref{eqn:Trule}) to momentum integral in (\ref{eqn:appV1}) to implement the finite temperature correction leads to a potential of the form presented in (\ref{eqn:fullV1eff}):
\begin{equation}
V_{eff}(\phi,T)=V_0(\phi)+\frac{T}{2}\sum_{n=-\infty}^{\infty}\int\frac{d^3k}{(2\pi)^3}\log(\omega_n^2+\vec{k}^2+V_0''(\phi)). 
\end{equation}
\subsection{Gauge degrees of freedom in Landau gauge}
Recalling the Goldstone equivalence theorem of gauge theory, one might doubt the necessity of counting the longitudinal polarization of a (massive) gauge field and its associated Goldstone mode as independent degrees of freedom when computing the effective potential in the  Landau gauge. Here we  clarify this fact in the simple case of an abelian Higgs model. To do so, we explicitly compute the one-loop contributions of the $U(1)$-gauge, ghost and Goldstone fields to the Higgs potential in the $R_\xi$ gauge. The effective potential turns out to be gauge-dependent, however there is no need to worry since it is not a physical observable. We work at $T=0$ but the following discussion can be driven the same way when the temperature is turned on, since we never evaluate momentum integrals. 

We begin with the gauge field ($A_\mu$). It will affect the Higgs potential at one-loop through the following term:
\begin{equation}
\Delta V_1^A(\phi)=-\frac{i}{2}\int\frac{d^4k}{(2\pi)^4}\log \mbox{det}\left(-i\widetilde{D}^{-1}_{\mu\nu}(k)\right), 
\end{equation}
where det acts on Lorentz indices. In the $R_\xi$ gauge the inverse propagator has the usual expression in momentum space:
\begin{equation}
-i\widetilde{D}^{-1}_{\mu\nu}(k)=\left(-k^2+m^2_A(\phi)\right)\Pi^T_{\mu\nu}(k)+\frac{1}{\xi}\left(-k^2+\xi m^2_A(\phi)\right)\Pi^L_{\mu\nu}(k)
\end{equation}
with $\Pi^T_{\mu\nu}(k)=\eta_{\mu\nu}-k_\mu k_\nu/k^2$ and $\Pi^L_{\mu\nu}(k)=k_\mu k_\nu/k^2$ being the transverse and longitudinal projectors respectively. Since the traces of $\Pi^{T,L}=3,1$ and the determinant are invariants, we can move to a basis where the matrices $\hat{\Pi}=C\Pi C^{-1}$ are diagonal and read:
\begin{equation}
\hat{\Pi}^T=\mbox{diag}(0,1,1,1)\quad ,\quad \hat{\Pi}^L=\mbox{diag}(1,0,0,0)
\end{equation}
In this basis the determinant can be easily evaluated and gives:
\begin{equation}
\Delta V_1^{A}(\phi)=-\frac{i}{2}\int\frac{d^4k}{(2\pi)^4}\left[3\log\left(-k^2+m^2_A(\phi)\right)+\log\left(-k^2+\xi m^2_A(\phi)\right)+\log\xi\right]\label{eqn:appV1A}
\end{equation}
Now we move to the Goldstone boson ($\chi$) and ghost contributions which are 
\begin{equation}
\Delta V_1^{\chi+ghost}(\phi)=-\frac{i}{2}\int\frac{d^4k}{(2\pi)^4}\log\left(-i\widetilde{D}^{-1}_\chi(k)\right)+i \int\frac{d^4k}{(2\pi)^4}\log\left(-i\widetilde{D}^{-1}_{ghost}(k)\right).
\end{equation}
Given that, in the abelian Higgs model, the inverse propagators are
\begin{eqnarray}
-i\widetilde{D}^{-1}_\chi(k)&=&k^2-m_\chi^2(\phi)-\xi m^2_A(\phi),\\
-i\widetilde{D}^{-1}_{ghost}(k)&=&k^2-\xi m^2_A(\phi),
\end{eqnarray}
with $m_\chi$ the mass the Goldstone receives from its Higgs couplings, we obtain
\begin{equation}
\Delta V_1^{\chi+ghost}(\phi)=\frac{i}{2}\int\frac{d^4k}{(2\pi)^4}\left[\log\left(-k^2+\xi m^2_A(\phi)\right)+i\pi-\log\left(1+\frac{m_\chi^2(\phi)}{-k^2+\xi m^2_A(\phi)}\right)\right].\label{eqn:appV1Gg}
\end{equation}
Gathering (\ref{eqn:appV1A}) and (\ref{eqn:appV1Gg}) together, we see the first terms of each expression cancel out, leaving only (in euclidean space)
\begin{equation}\label{eqn:appV1AGg}
\Delta V_1^{A+\chi+ghost}(\phi)=\frac{1}{2}\int\frac{d^4k_E}{(2\pi)^4}\left[3\log\left(k_E^2+m^2_A(\phi)\right)+\log\left(1+\frac{m_\chi^2(\phi)}{k^2_E+\xi m^2_A(\phi)}\right)\right]+\cdots,
\end{equation}
where $\cdots$ stand for constant terms  irrelevant for the potential. Taking $\xi=0$ to move to the Landau gauge, the last expression reduces to (up to an infinite constant)
\begin{equation}
\Delta V_{1,\xi=0}^{A+\chi+ghost}(\phi)=\frac{1}{2}\int\frac{d^4k_E}{(2\pi)^4}\left[3\log\left(k_E^2+m^2_A(\phi)\right)+\log\left(k^2_E+m^2_\chi(\phi)\right)\right]+\cdots,
\end{equation}
from which one clearly sees that, in this gauge, the factor of 3 for the massive gauge field is not altered by the addition of the Goldstone contribution. 

Another physically meaningful fixing choice is the unitary gauge $\xi\rightarrow\infty$. Sending the gauge fixing parameter to infinity in (\ref{eqn:appV1AGg}) implies the decoupling of the Goldstone contribution, as it should:
\begin{equation}
\Delta V_{1,\xi\rightarrow\infty}^{A+\chi+ghost}(\phi)=\frac{1}{2}\int\frac{d^4k_E}{(2\pi)^4}\left[3\log\left(k_E^2+m^2_A(\phi)\right)\right]+\cdots,
\end{equation}
Again the degrees of freedom of the gauge field are still 3 in this gauge.

From this discussion we see that   interpreting the factors in front of the $\log$ as the number of polarization states for the corresponding field  is only (accidentally) true in both $\xi=0,\infty$ gauges. Indeed, if one takes for instance the 't Hooft gauge  ($\xi=1$) the results are
\begin{equation}
\Delta V_{1,\xi=1}^{A+\chi+ghost}(\phi)=\frac{1}{2}\int\frac{d^4k_E}{(2\pi)^4}\left[2\log\left(k_E^2+m^2_A(\phi)\right)+\log\left(k^2_E+m^2_\chi(\phi)+m^2_A(\phi)\right)\right]+\cdots,
\end{equation}
where now the ``degrees of freedom" of $A_\mu$  reduce from 3 to 2 by this gauge choice. 

\subsection{Cancelation of  imaginary parts at small temperature}
\label{app:ImCancelation}

In Section~\ref{sec:ImCancelation}, we have shown that in the high temperature limit, a cancelation occurs between imaginary parts of the one-loop potential and the ring corrections. Here we want to show that this cancellation occurs also for smaller temperatures of order $T\sim |m_i(\phi)|$. Indeed by working out the integrals of (\ref{eqn:potTcorrections}) in order to isolate its imaginary part, we get
\begin{eqnarray}
\Im m\left[\Delta V_1^{T}(\phi,T)\right]
=
\sum_{i=h,\chi}\Theta(-m^2_i(\phi))\frac{n_iT^4}{4\pi^2}\int_0^{\frac{|m_i(\phi)|}{T}}dx x^2\left((4n+1)\pi-\sqrt{\frac{|m_i(\phi)|^2}{T^2}-x^2}\right),\nonumber
\end{eqnarray}
where $n$ is a positive integer which ensures that
\begin{equation}\label{eqn:Psheet}
-\pi<\frac{1}{2}\left(\pi-\sqrt{\frac{|m_i^2(\phi)|}{T^2}-x^2}\right)+2n\pi \leqslant \pi
\end{equation}
so that one stays on the principal sheet when taking the imaginary part of the logarithm,
whose branch is assumed to lie on the negative real axis of the complex plane. We can easily show that $n=0$ as long as $T>T_\pi\equiv |m_i(\phi)|/3\pi$, in which case the imaginary part becomes
\begin{eqnarray}
\Im m\left[\Delta V_1^{T}(\phi,T)\right]&=&\sum_{i=h,\chi}\Theta(-m^2_i(\phi))\frac{n_iT^4}{4\pi^2}\int_0^{\frac{|m_i(\phi)|}{T}}dx x^2\left(\pi-\sqrt{\frac{|m_i(\phi)|^2}{T^2}-x^2}\right),\nonumber\\
&=&\sum_{i=h,\chi}\Theta(-m^2_i(\phi))n_i\left[-\frac{|m_i(\phi)|^4}{64\pi}+\frac{|m_i(\phi)|^3 T}{12\pi}\right]\label{eqn:IMpartT}
\end{eqnarray}
and reproduces the same cancellation with (\ref{eqn:IMpartRING}) and (\ref{eqn:IMpart0}) as in the high temperature regime. 

For completeness we now consider the case of very low temperatures. As the temperature cools down below $T_\pi$, one begins needing to shift the imaginary part of the $\log$ by multiples of $2\pi$ to remain on the principal sheet of the complex plane. Furthermore, from (\ref{eqn:Psheet}) we see that 
\begin{equation}
 -2 \pi<
 (4n+1) \pi - \sqrt{\frac{|m_i(\phi)|^2}{T^2}-x^2} 
 \leq 2\pi.
\end{equation}
Thus 
\begin{equation}
-\sum_{i=h,\chi}\Theta(-m^2_i(\phi))n_i\frac{|m_i(\phi)|^3 T}{6\pi} 
<
\Im m\left[\Delta V_1^{T}(\phi,T)\right]
\leq
\sum_{i=h,\chi}\Theta(-m^2_i(\phi))n_i\frac{|m_i(\phi)|^3 T}{6\pi} 
\end{equation}
and we conclude that $\Im m\left[\Delta V_1^{T}(\phi,T)\right]$
vanishes as $T$ goes to zero, as it should.

%%%%%%%%%%%%%%%%%%%%%%%%%%%%%%%%%%%%%%%%%%%%%%%%%%%%%%%%%%%%%%%%%%%%%%%%%%%%%%%%%%%%%%%%%%%%%%%%%%%%ù

%%%%%%%%%%%%%%%%%%%%%%%%%%%%%%%%%%%%%%%%%%%%%%ù

\end{document}